\documentclass[%
 aip,
 amsmath,amssymb,
 reprint,%
]{revtex4-1}

\usepackage{graphicx}
\usepackage{dcolumn}
\usepackage{bm}

\usepackage[utf8]{inputenc}
\usepackage[T1]{fontenc}
\usepackage{mathptmx}

\begin{document}

\preprint{AIP/123-QED}

\title[ALE-LB for compressible flows]{Arbitrary Lagrangian-Eulerian formulation of lattice Boltzmann model for compressible flows on unstructured moving meshes}

\author{M. H. Saadat}
 \author{I. V. Karlin}%
 \email{karlin@lav.mavt.ethz.ch}
\affiliation{ 
Department of Mechanical and Process Engineering, ETH Zurich, 8092 Zurich, Switzerland
}%

\date{\today}

\begin{abstract}
We propose the application of the arbitrary Lagrangian-Eulerian (ALE) technique to a compressible lattice Boltzmann model for the simulation of moving boundary problems on unstructured meshes. To that end, the kinetic equations are mapped from a moving physical domain into a fixed computational domain. The resulting equations in the computational domain are then numerically solved using the second-order accurate finite element reconstruction on an unstructured mesh. It is shown that the problem regarding the geometric conservation law (GCL), which needs a special treatment in the ALE Navier-Stokes solvers, does not appear here and the model satisfies the GCL exactly. The model is validated with sets of simulations including uniform flow preservation and compressible flow past airfoil with plunging and pitching motions at different Mach numbers. It is demonstrated that the results are in good agreement with the experimental and other available numerical results in the literature. Finally, in order to show the capability of the proposed solver in simulating high-speed flows, transonic flow over pitching airfoil is investigated. It is shown that the proposed model is able to capture the complex characteristics of this flow which involves multiple weak shock waves interacting with the boundary and shear layers.  
\end{abstract}

\maketitle

\section{Introduction}
In recent years, there is a growing interest in studying both numerically \citep{bose2018investigating} and experimentally \citep{mackowski2015direct}, fluid flows in moving and deforming geometries in many physical phenomena and novel engineering applications. For example, in flapping flights of birds and insects, it is the motion of aerodynamic surfaces that produces thrust for forward motion and sustainable lift for airborne; or in marine animals such flapping motion generates propulsive and manoeuvring forces \citep{young2014review}. Understanding the underlying aerodynamics of these phenomena provides researchers with valuable insight into the origin of flight and its subsequent evolution in different species \citep{dyke2013aerodynamic}. Moreover, these natural phenomena have been a rich source of inspiration in design of engineering devices such as robotic devices, micro-air vehicles or in novel turbines that extract energy form wind and tidal waves using flapping foil motion. The flapping foil turbine concept is promising in turbine technologies as it is expected to be more efficient than vertical- and horizontal-axis turbines \citep{boudreau2018experimental} (for a review see \citet{young2014review, xiao2014review}). Other applications of flows with moving geometries appear in fluid-solid interactions (FSI) and rotor-stator flows to name a few.

Development of accurate and efficient numerical schemes for the simulation of fluid flows in complex domains remains a highly active research area in computational fluid dynamics (CFD). 
Presence of moving/deforming geometries in a flow adds another level of complexity to the computations, as one requires the numerical scheme not only to be able to handle moving domains, but to maintain accuracy and efficiency \citep{nguyen2010arbitrary}. From a physical point of view, moving domain problems usually involve vortex dominated unsteady flow with turbulence, separation and reattachment of the boundary layer. Numerically, capturing such a complex physics requires the numerical scheme to be accurate with small numerical dissipation.

The Lattice Boltzmann method (LBM) is a kinetic theory approach to CFD which has been proven as an accurate and reliable tool for the simulation of complex fluid flows ranging from turbulence \citep{chikatamarla2010lattice} and multiphase \citep{kim2015lattice} to micro-scale flows \citep{shan2006kinetic} and compressible flows \citep{frapolli2016entropic} (for a review on the application of LBM for complex fluid flows, see \citet{aidun2010lattice}). In the LBM, populations $f_i(\bm{x},t)$ associated with a set of discrete velocities $\mathcal{C} = \{\bm{c}_i,i=0,...,Q-1\}$  are  designed  to recover the target equations of continuum mechanics in the hydrodynamic limit. The evolution of populations is based on simple rules of propagation along the discrete velocities $\mathcal{C}$, and relaxation to a local equilibrium. This makes the LBM a simple and efficient alternative for conventional CFD solvers and an attractive candidate for simulation of flows with moving geometries. 

For handling moving complex geometries, most of the existing LB realizations employ a fixed background regular Cartesian grid which cuts the immersed moving object. Imposing no-slip boundary condition on the moving object is then achieved either through adding a force term into the equations \citep{feng2004immersed, chen2018immersed} or by replacing missing populations with some suitable approximation like non-equilibrium extrapolation \citep{guo2002extrapolation} or Grad's approximation \citep{dorschner2015grad}. 

Another approach which has widely been used in the Navier-Stokes (NS) framework for simulating moving domain problems is based on the so-called arbitrary Lagrangian-Eulerian (ALE) method \citep{hirt1974arbitrary, persson2009discontinuous, berndt2011two, boscheri2017arbitrary, jin2018ale}. In this method, 
the governing equations in the physical domain which is moving in space and time, are mapped into a fixed computational domain and then the resulting transformed equations are solved numerically \citep{persson2009discontinuous}. ALE method thus gives flexibility in handling moving domain problems as the physical mesh can move with arbitrary velocity independent of the fluid velocity \citep{boscheri2017arbitrary}. Another advantage of the ALE method is that it can handle moving/deforming domains with body-fitted mesh which is of crucial importance for high-Reynolds number flow simulations where small scale structures need to be resolved accurately. For problems involving rigid motion of objects and also problems with small deformation, it is possible to derive an analytical formula for the mapping function between physical and computational domains and that, in turn, greatly simplifies the computations. However, in problems with very large deformation, mapping function can become highly nontrivial and re-meshing might be required, which is computationally expensive and can easily make the simulation unfeasible. 

While numerous studies have been done on the ALE-NS solvers, limited number of works can be found in the literature about applying ALE method in the context of LBM. Noteworthy is the model proposed by \citet{meldi2013arbitrary}, which is based on the combination of ALE and over-set grid methods and was used for simulation of low-speed incompressible flow. 

The aim of this paper is to investigate the application of the ALE method in LB framework for the simulation of moving domain problems. It is well known that most of the conventional LB models in the literature are limited to low-speed isothermal incompressible flows. That is due to the insufficient isotropy and lack of Galilean invariance of the standard lattices ($D2Q9$ in two dimensions and $D3Q27$ in three dimensions, where $DdQn$ model refers to $d$ dimension model with $n$ discrete velocities). A systematic way to overcome this severe limitation could be achieved by increasing the number of discrete velocities and use the hierarchy of higher-order (multi-speed) lattices \citep{chikatamarla2009lattices, frapolli2016entropic}. However, employing high-order lattices comes at the price of increasing the computational cost. Another recent approach, which maintains the simplicity and efficiency of the standard lattices, is to introduce appropriate correction terms into the kinetic equations in order to compensate the error terms resulting from the low symmetry of the standard lattices (see e.g. the models proposed by \citet{prasianakis2008lattice}, \citet{feng2015three} and \citet{ huang2019lattice}). Following this approach, we have recently introduced a compressible LB model on standard lattices which can recover the full Navier-Stokes-Fourier (NSF) equations with adjustable Prandtl number and adiabatic exponent in the hydrodynamic limit \citep{saadat2019lattice}. We then enriched our model by employing the concept of the shifted lattices \citep{frapolli2016lattice} and showed that the model works pretty well for compressible flows up to moderate supersonic regime with shock waves. 

In this paper, we apply the ALE method to the compressible LB model \citep{saadat2019lattice} as it gives us a unified flow solver which covers subsonic to moderately supersonic regimes. To the best of our knowledge, LBM has not been investigated for the simulation of compressible flows with moving bodies. It should, however, be emphasized that the ALE formulation given below is general and, in principle, can be applied to any lattice kinetic model, including incompressible \citet{mattila2017high,bosch2015entropic}, thermal \citet{karlin2013consistent} or compressible models \citet{frapolli2016entropic}.

Since transformed equations should be solved in the ALE method, exact propagation on space-filling lattice is not possible anymore and an off-lattice scheme should be adopted. Among various off-LB schemes existing in the literature, such as finite difference (FD) or finite volume (FV) LB schemes \citep{zarghami2014finite}, the semi-Lagrangian scheme based on the finite-element interpolation \citep{kramer2017semi} has been shown to be an efficient and accurate scheme which maintains the advantage of high Courant–Friedrichs–Lewy (CFL) number, while removes the restriction of using regular lattice. The semi-Lagrangian finite element scheme makes it also possible to employ unstructured body-fitted grid which gives us more flexibility in handling complex geometries and is more efficient in flows with high-Reynolds number. The application of the semi-Lagrangian scheme with finite element interpolation has been studied for incompressible LB models in both laminar \citep{kramer2017semi} and turbulent regimes \citep{di2018simulation}.

The rest of the paper is organized as follows: The ALE formulation of the LB model is discussed in Sec.\ref{Sec:ModelDiscription}. The detailed numerical implementation of the ALE-LB model on unstructured mesh is presented in Sec.\ref{Sec:NumericalImple}. It is also shown that the ALE-LB model satisfies the geometric conservation law (GCL) exactly. The kinetic equations of the compressible LB model along with the implementation of no-slip wall boundary conditions are briefly reviewed in Sec.\ref{Sec:Kinetic_eqns}. In  Sec.\ref{Sec:Results}, the model is validated through simulation of benchmark test-cases, including free-stream preservation and compressible flow over NACA0012 airfoil in plunging and pitching motions. Moreover, in order to test the model's performance in simulating high-speed flows, the transonic flow over pitching airfoil is considered in this section. Finally, some conclusions are drawn in Sec.\ref{Sec:conclusion}.

\section{ALE formulation of the lattice Boltzmann} \label{Sec:ModelDiscription}
Consider the Boltzmann equation in a physical domain $(\bm{x},t)$
\begin{align}
    \frac{\partial f_i}{\partial t} + \bm{c}_i . \bm{\nabla_x } f_i = \Omega_i, \label{eq:Boltzmann}
\end{align}
where $f_i(\bm{x},t)$ are populations of discrete velocities $\bm{c}_i,i=0,...,Q-1$, and $\Omega_i$ is the collision operator. The goal here is to transform Eq. (\ref{eq:Boltzmann}) from a physical domain $(\bm{x},t)$ to a fixed computational domain $\left( \bm{X},t_0 \right)$. 

We assume that there exists a continuous time dependent mapping between physical and computational domains, denoted by $G$, such that $\bm{x} = G \left(\bm{X},t \right)$ \citep{persson2009discontinuous}. The time derivative in Eq. (\ref{eq:Boltzmann}) can then be re-written as
\begin{align}
    \frac{\partial f_i}{\partial t} = \frac{d f_i}{d t} - \bm{V_G}.\bm{\nabla_X} f_i 
    = \frac{\partial f_i}{\partial t}|_{\bm{X}} - \bm{V_G}.\bm{\nabla_X} f_i, \label{eq:Time_Derivative}
\end{align}
where the time derivative $\frac{\partial f_i}{\partial t}|_{\bm{X}}$ is at constant $\bm{X}$, spatial derivatives are taken with respect to $\bm{X}$, and $\bm{V_G}$ denotes the mapping velocity as
\begin{align}
    \bm{V_G} = \frac{\partial G}{\partial t}|_{\bm{X}}.
\end{align}
Using a simple chain rule, the spatial terms in Eq. (\ref{eq:Boltzmann}) can also be written as
\begin{align}
    \bm{\nabla_x } f_i = \bm{g}^{-1} \bm{\nabla_X } f_i. \label{eq:Chain_Rule}
\end{align}
Here, $\bm{g}^{-1}$ is the inverse of the Jacobian matrix of mapping, which for two-dimensional problems can be computed as
\begin{align}
    \bm{g}^{-1} = \frac{1}{\text{det } \bm{g}}
\begin{bmatrix}
y_Y & -y_X \\ 
-x_Y & x_X 
\end{bmatrix},
\end{align}
where $x_X$, $x_Y$, $y_X$ and $y_Y$ are mapping metrics and
\begin{align}
    \text{det } \bm{g} = x_X y_Y - y_X x_Y, \label{eq:g_inv}
\end{align}
is the determinant of the Jacobian matrix.
Substituting Eq. (\ref{eq:Time_Derivative}) and Eq. (\ref{eq:Chain_Rule}) into Eq. (\ref{eq:Boltzmann})
\begin{align}
    \frac{\partial f_i}{\partial t}|_{\bm{X}} + \bm{c}_i . \bm{g}^{-1} \bm{\nabla_X } f_i -  \bm{V_G}.\bm{\nabla_X} f_i = \Omega_i,
\end{align}
which can be further simplified as
\begin{align}
    \frac{\partial f_i}{\partial t}|_{\bm{X}} + \left( \bm{g}^{-1,T} \bm{c}_i - \bm{V_G} \right) . \bm{\nabla_X } f_i = \Omega_i,
\end{align}
where the superscript $T$ denotes the transpose of a matrix. By defining transformed discrete velocities $\hat{\bm{c}}_i$ as
\begin{align}
\hat{\bm{c}}_i = \left( \bm{g}^{-1,T}\bm{c}_i - \bm{V_G} \right), \label{eq:Trasformed_Discrete_Vel}   
\end{align}
the Boltzmann equation in a fixed computational domain can be written in a simple form as
\begin{align}
    \frac{\partial f_i}{\partial t}|_{\bm{X}} + \hat{\bm{c}}_i . \bm{\nabla_X } f_i = \Omega_i. \label{eq:Boltzmann_fixed}
\end{align}
As it can be seen, the only difference between Eq. (\ref{eq:Boltzmann_fixed}) and Eq. (\ref{eq:Boltzmann}) is in discrete velocities. We can therefore conclude that, the ALE method is applicable to any lattice kinetic model just by using the transformed discrete velocities as defined in Eq. (\ref{eq:Trasformed_Discrete_Vel}). 

Now, Eq. (\ref{eq:Boltzmann_fixed}) can be discretized using conventional scheme used in the standard LB, i.e. through propagation and collision steps
\begin{align}
    {f_i}\left( {\bm{X},t } \right) - {f_i} \left( \bm{X} - \bm{\hat{\bm{c}}}_i\delta t,t - \delta t \right) =  \Omega_i. \label{eq:Boltzmann_Discretized}
\end{align}

It is evident from Eq. (\ref{eq:Trasformed_Discrete_Vel}) that the discrete velocities $\hat{\bm{c}}_i$ are not necessarily integer numbers anymore. Thus, unlike standard LB, exact propagation on space-filling lattice is not possible here and interpolation is required during the propagation step.
Here, we use a second-order accurate finite-element interpolation which has been shown to be accurate and less dissipative compared to other off-LB schemes \citep{kramer2017semi}. It also makes it possible to employ unstructured mesh which is more suitable in handling complex geometries.

\section{Numerical implementation} \label{Sec:NumericalImple}
The numerical implementation involves propagation and local collision (see Eq. (\ref{eq:Boltzmann_Discretized})). 
Performing propagation on unstructured mesh removes the restriction of the classical LBM related to the regular lattice and turns the propagation problem into an interpolation one.

\begin{figure}
    \centering
    \includegraphics[width=0.5\textwidth]{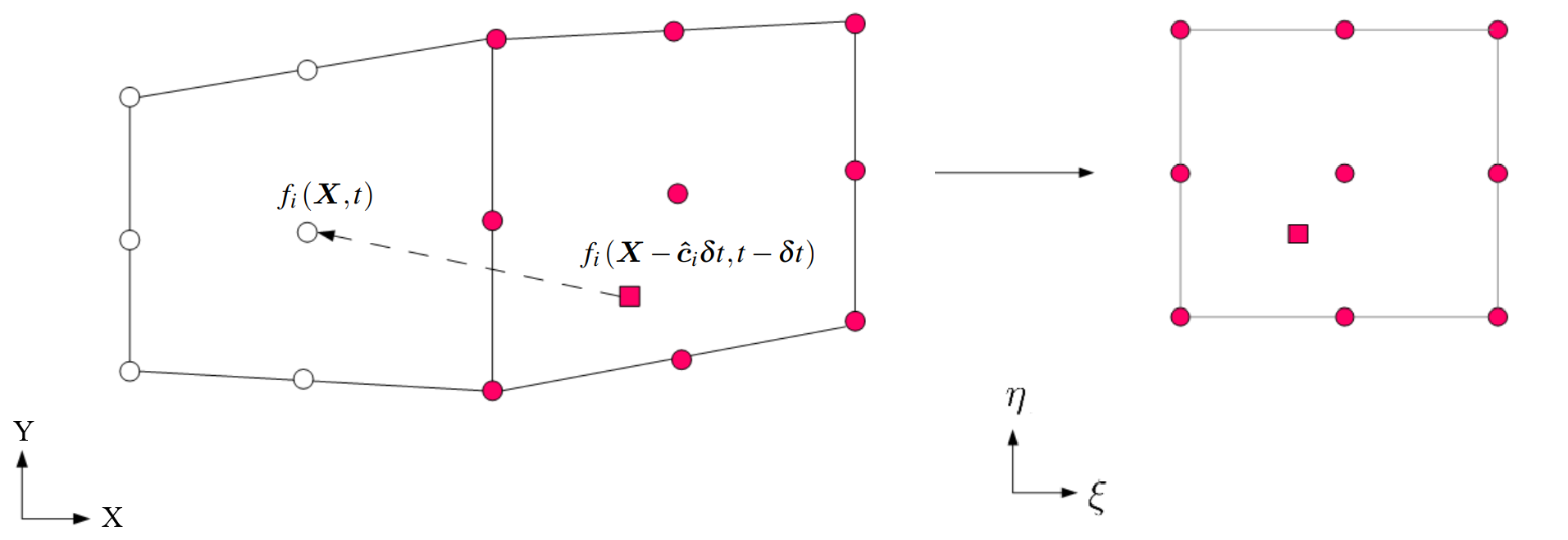}
    \caption{Schematic of a second-order finite element mesh, the propagation along the discrete velocity $\bm{\hat{c}}_i$ and mapping from the global coordinate $(X,Y)$ to local coordinate $(\xi, \eta)$.}
    \label{fig:FE_Mesh}
\end{figure}

The computational domain is first discretized into an irregular mesh. Then in order to perform propagation step at each grid node, similar to standard LB, we follow the characteristics curve of the LB equation backward in time to find the departure point of each grid node. Populations $f_i$ at the corresponding departure point of each grid node $\bm{X}-\delta t \hat{\bm{c}}_i$ are reconstructed through using the second-order accurate finite-element interpolation scheme \citep{kramer2017semi}. In two dimensions, the interpolation is based on the second-order Lagrange polynomials defined on nine equidistant collocation nodes. An example of a propagation on a second-order finite element mesh is depicted in Fig. \ref{fig:FE_Mesh}.
The propagation step can, therefore, be written as
\begin{align} 
f_i \left(\bm{X} , t\right) &= f_i \left(\bm{X}-\bm{\hat{c}}_i\delta t,t-\delta t\right)
= \sum_{s=1}^{9}N_s(\bm{\xi}_{dp})f_i(\bm{\xi}_s, t -\delta t), \label{eq:Propagation}
\end{align}
where $N_s(\bm{\xi}_{dp})$ denote the values of the shape functions \citep{zienkiewicz2005finite}, written in the local coordinate system $\bm{\xi} = (\xi,\eta)$, $ (-1 \le \xi,\eta \le 1$), at the departure point (red square in Fig. \ref{fig:FE_Mesh}), $f_i(\bm{\xi}_s, t -\delta t)$ are the values of populations $f_i$ at the collocation nodes (red circles in Fig. \ref{fig:FE_Mesh}), and $s=9$ is number of collocation points.   

Therefore, semi-Lagrangian propagation on unstructured finite-element mesh requires two steps:
First, computing the local coordinates of the departure point $\bm{\xi}_{dp}$ (see Fig. \ref{fig:FE_Mesh}). Here, a bi-linear transformation is used to transform the computational cells into a reference unit cell. Thus, finding the local coordinates requires solving a non-linear system of equations resulting from
\begin{align}
    \bm{X}_{dp} = \sum_{s=1}^4 N_s(\bm{\xi_{dp}}) \bm{X_s}. \label{eq:NonLinear_Eq}
\end{align}
Unlike stationary case, where the location of departure point for each node is constant during the simulation, in moving case, the departure point is moving and therefore, the non-linear equation (\ref{eq:NonLinear_Eq}) should be solved in each time step.
Second, computing the values of the populations at the departure point by means of the values of the populations at collocation nodes (red circles), i.e. by using Eq (\ref{eq:Propagation}).

After propagation, the post-collision populations are computed.

\subsection{Geometric conservation law}
The problem of geometric conservation law (GCL) was first introduced in \citet{thomas1979geometric}, where it was shown that the numerical discretization errors associated with mapping metrics can induce errors in the computed flow field which might lead to numerical instabilities \citep{thomas1979geometric}. This problem has been widely studied in the NS solvers and different strategies have been proposed for satisfying the GCL in that context \citep{persson2009discontinuous, deng2011geometric, abe2014geometric}.

In order to mathematically check the GCL, a uniform flow should satisfy Eq. (\ref{eq:Boltzmann_Discretized}). As the collision term vanishes with constant uniform flow, we just need to insert a constant solution $f_i(\bm{X},t)=f_i^0$ into Eq. (\ref{eq:Propagation})
\begin{align} 
f_i ^0 = \sum_{s=1}^{}N_s f_i^0. \label{eq:GCL1}
\end{align}
Since the summation of shape functions is, by construction, equal to one $\left( \sum_{s=1}^{}N_s = 1 \right)$, the RHS of Eq. (\ref{eq:GCL1}) is simplified as
\begin{align} \label{eq:}
 \sum_{s=1}^{}N_s f_i^0 = f_i^0\sum_{s=1}^{}N_s = f_i^0,
\end{align}
and, therefore, the present model satisfies the GCL exactly. 

Before proceeding to results section, we briefly describe the compressible LB model used in this study.

\section{Kinetic equations} \label{Sec:Kinetic_eqns}
The kinetic model used in this study is the two-population compressible LB model on standard lattices, which recovers the full NSF equations in the hydrodynamic limit \citep{saadat2019lattice}. 

The kinetic equations of this model written in the computational domain are as follows
\begin{align}{}
	{f_i}\left( {\bm{X},t } \right) - {f_i}(\bm{X} - \bm{\hat{c}}_i\delta t,t - \delta t) &=  \omega (f_i^{eq}-{f_i}) + \delta t \phi_i, \label{eq:feqn} \\
	\begin{split}
	{g_i}\left( {\bm{X},t } \right) - {g_i}(\bm{X} - \bm{\hat{c}}_i\delta t,t - \delta t) &=
	\omega (g_i^{eq} - g_i) \\&+ ({\omega _1} - \omega)(g_i^* - {g_i}),   \label{eq:geqn}
	\end{split}
\end{align}
where $\bm{\hat{c}}_i$ are transformed discrete velocities computed using Eq. (\ref{eq:Trasformed_Discrete_Vel})
\begin{align}
\hat{\bm{c}}_i = \left( \bm{g}^{-1,T}\bm{c}_i - \bm{V_G} \right),     
\end{align}
and standard set of discrete velocities $\bm{c}_i$ in two dimensions and with $Q=9$ ($D2Q9$ model) is defined as
\begin{align}
\bm{c}_i = \left( {{c_{ix}},{c_{iy}}} \right)^T, \text{	} i = 0,...,Q-1; \text{	} {c_{i\alpha }} \in \left\{ { - 1,0, + 1} \right\}, \text{ } \alpha = x,y.  
\end{align}
The $\phi_i$ terms in Eq.(\ref{eq:feqn}) are correction terms responsible for canceling out the spurious terms in the momentum equation, resulting from low symmetry of the standard lattices (for detail of deriving correction terms see \citet{saadat2019lattice}), $g_i^*$ is a quasi-equilibrium population, and $f_i^{eq}$, $g_i^{eq}$ are local equilibria which satisfy the local conservation laws for the density $\rho$, momentum $\rho \bm{u}$ and total energy $\rho E$,
\begin{alignat}{3}
    \sum_{i=0}^{Q-1} \{ 1, \bm{c}_i \} f_i &= \sum_{i=0}^{Q-1} \{ 1, \bm{c}_i \} f_i^{eq} &&= \{ \rho, \rho \bm{u} \},   \\
    \sum_{i=0}^{Q-1}  g_i &= \sum_{i=0}^{Q-1}  g_i^{eq} &&=  2\rho E.
\end{alignat}
The temperature is defined by
\begin{equation}
	T = (1/{C_v})(E - {u^2}/2), \label{eq:Temp}
\end{equation}
where $C_v$ is the specific heat of ideal gas at constant volume. The relaxation parameters $\omega$ and ${\omega _1}$ are related to the dynamic viscosity $\mu$ and thermal conductivity $\kappa$
\begin{align}
	\mu &= \left( {\frac{1}{\omega } - \frac{1}{2}} \right) \rho T \delta t, \\
	\kappa &= C_p \left( {\frac{1}{\omega_1 } - \frac{1}{2}} \right) \rho T \delta t.
\end{align}
Below, a system of units is used where the universal gas constant is set to one, $R=1$. Consequently, $C_p = C_v + 1$ is the specific heat at constant pressure and the Prandtl number is Pr $=C_p\mu / \kappa$; $\gamma = C_p / C_v$ is the adiabatic exponent which can be freely adjusted.

The equilibrium $f$- populations can be written in a product-form as
\begin{equation}
		f_i^{eq} = \rho {\Phi _{c_{ix}}}{\Phi _{c_{iy}}},  \label{eq:feq}
		\end{equation}
		where
		\begin{align}
		{\Phi _{ - 1}} &= \frac{{ - {u_\alpha } + u_\alpha ^2 + T}}{2}, \\
		{\Phi _{0}} &= 1 - \left( {u_\alpha ^2 + T} \right), \\
		{\Phi _{ + 1}} &= \frac{{{u_\alpha  } + u_\alpha  ^2 + T}}{2},
\end{align}

The populations $g_i^{eq}$, $g_i^{*}$ are constructed using the following general form
\begin{align}
            G_i &= W_i \left(  M_0 + \frac{{M_\alpha {c_{i\alpha }}}}{T} \right.  
		    \left. + \frac{{(M_{\alpha \beta } - M_0 T{\delta _{\alpha \beta }})({c_{i\alpha }}{c_{i\beta }} - T{\delta _{\alpha \beta }})}}{{2{T^2}}} \right) + \Psi_i, \label{eq:geq_gstar}
\end{align}
where ${W_i}$ are temperature-dependent weights
	\begin{equation}
	{W_i} = {W_{{c_{ix}}}}{W_{{c_{iy}}}}, \label{eq:Wi}
	\end{equation}
	with 
	\begin{align}
	{W_{ - 1}} &= \frac{{ T}}{2}, \\
	{W_{0}} &= 1 - T, \\
	{W_{ + 1}} &= \frac{{ T}}{2},
	\end{align}
and other terms required for the computations are provided in Table \ref{tab:Eq_Moments} and defined as
\begin{table} \centering
\caption{Moments needed for the computation of $g_i^{eq}$ and $g_i^{*}$. }
\label{tab:Eq_Moments}
\begin{tabular}{ccccl}
\hline \hline
$G_i$           & $M_0$           & $M_\alpha$      & $M_{\alpha \beta}$    & $\Psi_i$  \\ \hline
$g_i^{eq}$      & $2 \rho E$     & $q_\alpha^{eq}$       & $R_{\alpha \beta}^{eq}$  & $\psi_i$ \\ 
$g_i^{*}$       & $2 \rho E$     & $q^{*}_\alpha$   & $R_{\alpha \beta}^{eq}$ & $\psi_i$ \\ \hline \hline
\end{tabular}
\end{table}
\begin{alignat}{2}
	q_\alpha ^{eq} &= \sum\limits_{i = 0}^{Q-1} {{c_{i\alpha }}} g_i^{eq} = 2\rho {u_\alpha }(E+T),   \\
	\begin{split}
	R_{\alpha \beta }^{eq} &= \sum\limits_{i = 0}^{Q-1} {{c_{i\alpha }}{c_{i\beta }}} g_i^{eq} = 
	2\rho E(T{\delta _{\alpha \beta }} + {u_\alpha }{u_\beta }) 
	 \nonumber \\& + 2\rho T(T{\delta _{\alpha \beta }} + 2{u_\alpha }{u_\beta }),
	\end{split} \\
	q_\alpha ^{*} &= \sum\limits_{i = 0}^{Q-1} {{c_{i\alpha }}} g_i^{*} = 2{u_\beta }({P_{\alpha \beta }} - P_{\alpha \beta }^{eq}), \\
	P_{\alpha \beta }^{eq} &= \sum\limits_{i = 0}^{Q-1} {{c_{i\alpha }}} {c_{i\beta }}f_i^{eq} = \rho {u_\alpha }{u_\beta } + \rho T{\delta _{\alpha \beta }}, \\
	\psi_i &= B_{i \alpha} \left( \rho \left(1 - 3T\right)\left({T^2} + 2{u_\alpha^2}T + E{u_\alpha^2}\right)/T \right), \\
    {B_{i\alpha }} &= \left\{ \begin{array}{l}
		1, {\rm{ for \text{	} }} \boldsymbol{c}_i = \boldsymbol{0},\\ \nonumber
		-\left| {{c_{i\alpha }} - \frac{1}{2}{c_{i\alpha }}c_i^2} \right| , {\rm{ otherwise}}{\rm{.}}
		\end{array} \right. \label{eq:i} 
\end{alignat}
Note that, summation convention is used in above equations.

Finally, the correction terms $\phi_i$ can be computed as
\begin{align}
		\phi_i = A_{i\alpha } X_\alpha, 
\end{align}
where 
\begin{align}
        {X_\alpha } &=  - {\partial _\beta }\left[ {\left( \frac{\mu}{\rho T} \right){\partial _\gamma }{{Q'}_{\alpha \beta \gamma }}} \right], \label{eq:Xa} \\
		{A_{i\alpha }} &= {c_{i\alpha }} - \frac{1}{2}{c_{i\alpha }}c_i^2,  \label{eq:Ai}
\end{align}
and 
${Q'}_{\alpha \beta \gamma }$ is the deviation of the the third-order equilibrium moment from the continuous Maxwell-Boltzmann moment (for further detail see \citet{saadat2019lattice})
\begin{align} \label{eq:DeviationTerms}
\small
{Q'}_{\alpha \beta \gamma } = 
    \left\{\begin{matrix}
\rho u_\alpha (1 - 3T) - \rho u_\alpha ^3, \text{ if }\alpha = \beta = \gamma,  \\ 
0, {\text{ if }\alpha  \ne \beta ,{\text{ or }}\alpha  \ne \gamma ,{\text{ or }}\beta  \ne \gamma. }\\
\end{matrix}\right.
\end{align}

It is important to note that, spatial derivatives in the correction terms $\phi_i$ (Eq.(\ref{eq:Xa})) should also be transformed from the physical to computational domain. This can be done using a simple chain rule similar to Eq. (\ref{eq:Chain_Rule}). For a generic variable $K$, we can write
\begin{align}
    \partial_{\bm{x}}  K = \bm{g}^{-1} \partial_{\bm{X}}  K. \label{eq:Ph_Derivative}
\end{align}
Here, $\bm{g}^{-1}$ is computed using Eq. (\ref{eq:g_inv}), and
\begin{align}
    \partial_{\bm{X}} K = \bm{J}^{-1} \sum_{s}^{} K_s \partial_{\bm{\xi}} N_s, \label{eq:FE_Derivative}
\end{align}
where $\bm{J}^{-1}$ is the inverse of the Jacobian matrix of transformation of a computational cell to a unit cell computed with
\begin{align}
    \bm{J}^{-1} = \frac{1}{\text{det } \bm{J}}
\begin{bmatrix}
\partial_{\eta} y & -\partial_{\xi} y \\ 
-\partial_{\eta} x & \partial_{\xi} x 
\end{bmatrix},
\end{align}
and
\begin{align}
    \text{det } \bm{J} = \partial_{\xi} x   \partial_{\eta} y - \partial_{\xi} y \partial_{\eta} x,
\end{align}
is the determinant of the Jacobian matrix of transformation. The metrics of transformation $\partial_{\xi} x, \partial_{\eta} x, \partial_{\xi} y,\partial_{\eta} y$ are computed with the following formula
\begin{align}
    \partial_{\bm{\xi}} \bm{x} = \sum_{s}^{} \bm{x}_s \partial_{\bm{\xi}} N_s.
\end{align}
Note that, the nodes on the element edges are assigned to the element with the larger area.

\subsection{No-slip wall boundary condition}
The no-slip wall boundary condition (BC) used in this work is based on those proposed in \citet{dorschner2015grad}. The general idea is to replace the missing populations during the propagation step with the following expression 
\begin{align}
    f_i^{miss} &= f_i^{eq}(\rho_{tgt},\bm{u}_{tgt},T_{tgt}) 
    + \delta t f_i^{(1)}(\rho_{tgt},\bm{u}_{tgt},T_{tgt}, \triangledown \bm{u}_{tgt}, \triangledown T_{tgt} ), \\
    g_i^{miss} &= g_i^{eq}(\rho_{tgt},\bm{u}_{tgt},T_{tgt}) 
    + \delta t g_i^{(1)}(\rho_{tgt}, \bm{u}_{tgt}, T_{tgt}, , \triangledown \bm{u}_{tgt}, \triangledown T_{tgt}),
\end{align}
where $f_i^{eq}$, $g_i^{eq}$ are equilibrium parts computed from (\ref{eq:feq}) and (\ref{eq:geq_gstar}), $f_i^{(1)}$, $g_i^{(1)}$ are non-equilibrium parts and $\rho_{tgt}$, ${\boldsymbol{u}_{tgt}}$ and $T_{tgt}$ are target values which will be specified later. The non-equilibrium parts are obtained based on the Grad's approximation and using the general formula (\ref{eq:geq_gstar}) with the non-equilibrium moments given in Table \ref{tab:Neq_Moments} \citep{frapolli2016entropic}
\begin{table} \centering
\caption{Moments needed for the computation of $f_i^{(1)}$ and $g_i^{(1)}$. } 
\label{tab:Neq_Moments}
\begin{tabular}{ccccl}
\hline \hline
$G_i$           & $M_0$           & $M_\alpha$      & $M_{\alpha \beta}$  &$\Psi_i$ \\ \hline
$f_i^{(1)}$     & $0$             & $0$             & $P_{\alpha \beta}^{(1)}$  &$0$ \\ 
$g_i^{(1)}$     & $0$             & $q^{(1)}_\alpha$   & $R_{\alpha \beta}^{(1)}$ &$0$ \\ \hline \hline
\end{tabular}
\end{table}
\begin{align}
   P_{\alpha \beta}^{(1)} &= -\frac{1}{\omega} \rho T \left( S_{\alpha \beta} - \frac{1}{C_v}\partial_\gamma u_\gamma \delta_{\alpha \beta}    \right), \\
   q^{(1)}_\alpha &= -\frac{2}{\omega_1} \rho C_p T \partial_\alpha T + 2u_\beta P_{\alpha \beta}^{(1)}, \\
  R_{\alpha \beta}^{(1)} &= -\frac{2}{\omega_1} \rho T \left[S_{\alpha \beta} \left( E + 2T \right) + u_\alpha \partial_\beta E + u_\beta \partial_\alpha E \right],
\end{align}
where the strain rate tensor is
\begin{align}
    S_{\alpha \beta} = \partial_\alpha u_\beta + \partial_\beta u_\alpha.
\end{align}

\begin{figure}
    \centering
    \includegraphics[width=0.5\textwidth]{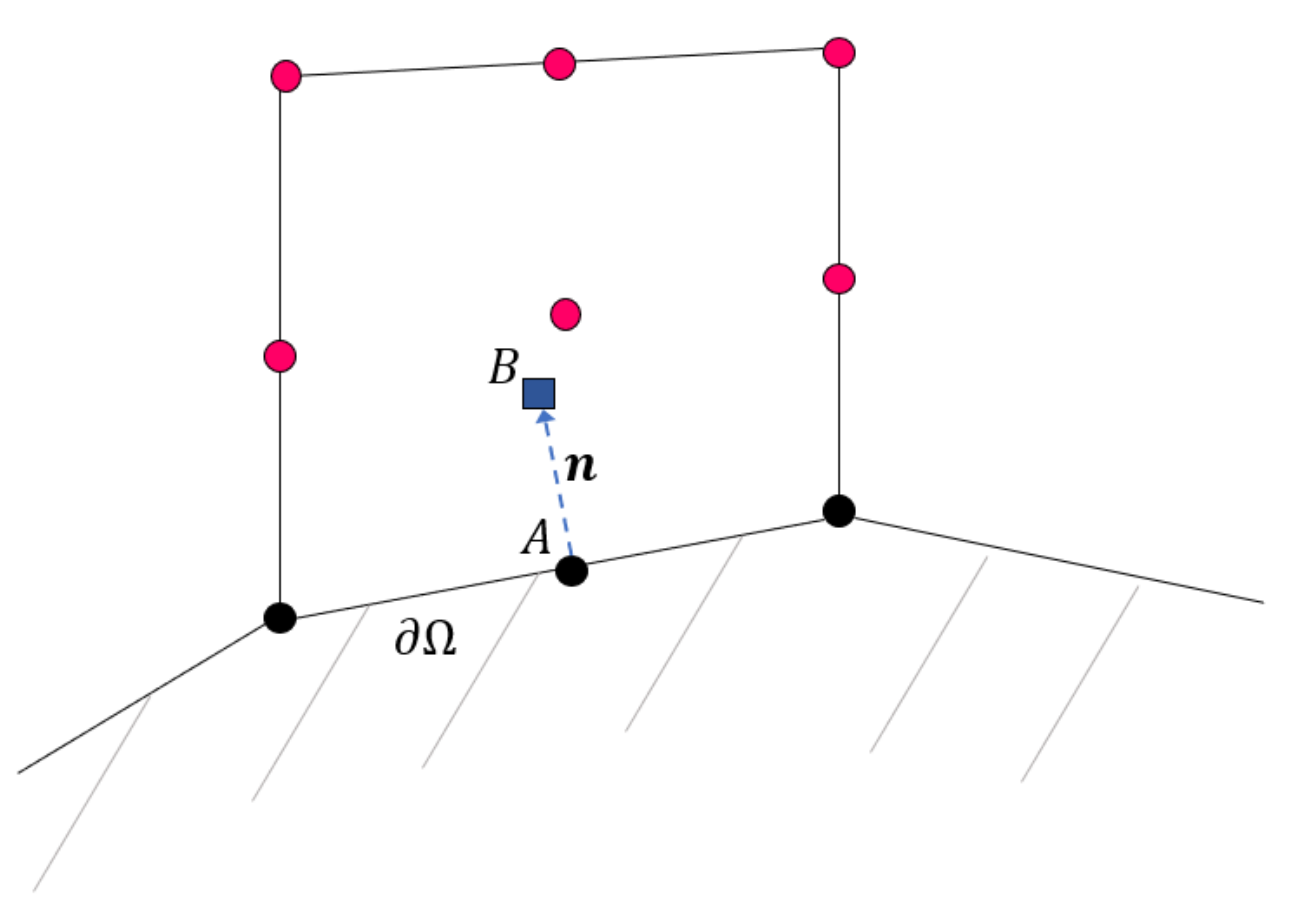}
    \caption{Schematic representation of the wall boundary condition implementation.}
    \label{fig:FE_BC}
\end{figure}

For computing target values, if missing populations belong to points on the wall (black circles in Fig. \ref{fig:FE_BC}), target velocities are wall velocities, $\bm{u}_{tgt} = \bm{u}_{wall}$ and target density and temperature (for adiabatic wall) are obtained by setting
\begin{align}
    \frac{\partial \rho}{\partial \bm{n}}|_{wall} = 0, \\
    \frac{\partial T}{\partial \bm{n}}|_{wall} = 0,
\end{align}
where $\bm{n}$ is the normal direction to the wall boundary $\partial \Omega$. Given the normal direction $\bm{n}$, its end point $B$ and considering the distance from $A$ to $B$ as $||\bm{n}||=\delta t$, the values of density and temperature at $B$ can be evaluated using a finite element interpolation
\begin{align}
    \rho_{B} = \sum_{s=1}^{9}N_s \rho_s,
\end{align}
where $N_s$ are shape functions and $\rho_s$ are the magnitude of density at nine collocation points (circles in Fig. \ref{fig:FE_BC}). Once $\rho_B$ is found, the first-order approximation for the normal derivative is assumed
\begin{align}
     \frac{\partial \rho}{\partial \bm{n}}|_{wall} = \frac{\rho_B - \rho_A}{||\bm{n}||} = 0.
\end{align}
Therefore, the target value can be approximated as
\begin{align}
    \rho_{tgt} = \rho_A = \rho_B.
\end{align}
The same procedure is applied for computing target temperature $T_{tgt}$.
Note that if missing populations belong to points which do not lie on the wall boundaries (red circles in Fig. \ref{fig:FE_BC}), the local quantities of the previous time step are used as target values.

The evaluation of spatial gradients in non-equilibrium parts is performed using (\ref{eq:Ph_Derivative}). It was demonstrated in \citet{dorschner2015grad} that the first-order accurate evaluation of spatial derivatives is sufficient. 

\section{Results} \label{Sec:Results}
In this section, the model presented above is tested numerically through simulation of benchmark cases for moving boundary problems. First, the GCL of the model is validated. Then, we investigate the flapping airfoil under pure plunging and pitching motions, which is relevant in many physical applications including the flight of small fliers or micro air vehicles. 
All simulations are performed with $\gamma = 1.4$, Pr $=0.71$, $D2Q9$ lattice model and the adiabatic wall assumption.

\subsection{Free-stream preservation}
\begin{figure}
    \centering
    \includegraphics[width=0.5\textwidth]{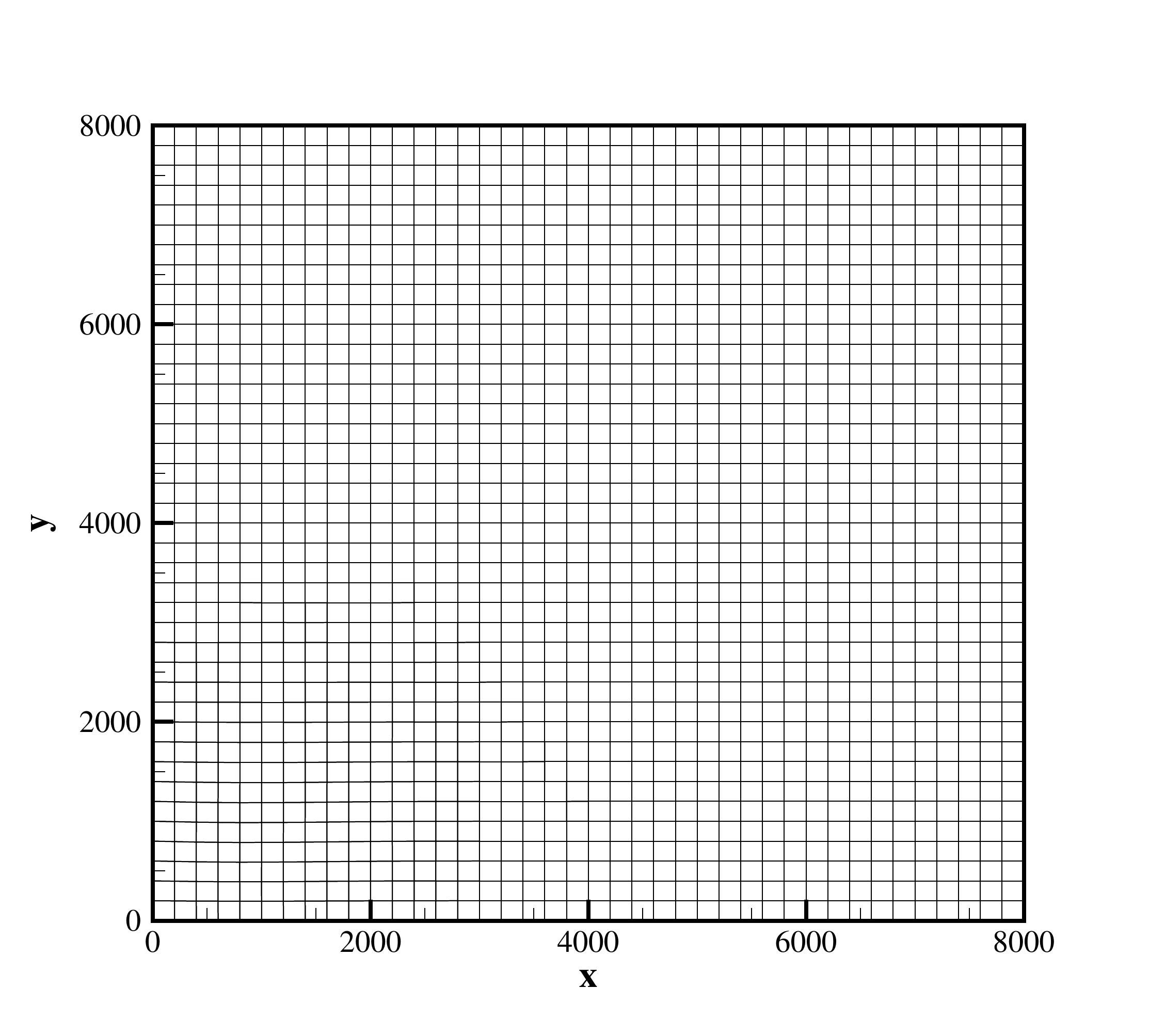}
    \includegraphics[width=0.5\textwidth]{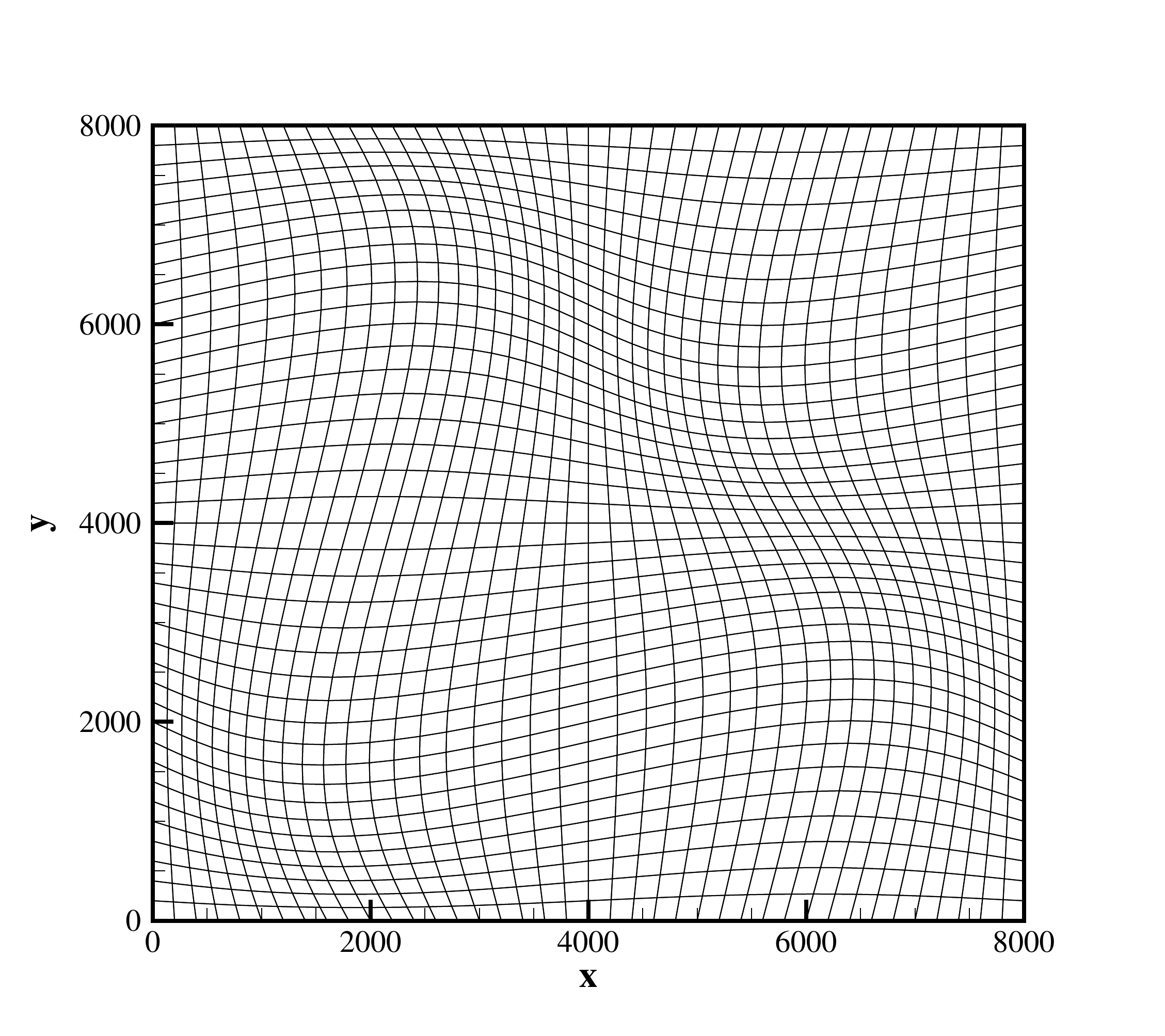}
    \caption{Motion of the mesh at non-dimensional time $t^{*} = 0$ (left) and $t^{*} = 1$ (right).}
    \label{fig:GCL_Mesh}
\end{figure}

The first test-case is to check the GCL of the model, i.e. to ensure the exact conservation of the free-stream condition under arbitrary movement of the mesh. We consider a uniform flow with $Ma = u_\infty /  \sqrt{\gamma T} = 0.2$ and $T = 0.2$ in a square domain of size $L = 8000$. The mesh motion is defined through the following mapping function
\begin{align}
    x(t) &= X + 500 sin(2 \pi X / L) sin ((2 \pi Y / L)) sin((2 \pi t / t_0)), \\
    y(t) &= Y + 500 sin(2 \pi X / L) sin ((2 \pi Y / L)) sin((2 \pi t / t_0)),
\end{align}
with the reference time $t_0 = 1.5 L/u_\infty$. Figure \ref{fig:GCL_Mesh} shows the mesh at two different non-dimensional times $t^{*} = t u_\infty /L$. We compute the solution until non-dimensional time $t^{*} = 1$ using three different uniform grids, and the relative errors $\epsilon = \sum{|u-u_\infty|}/\sum{|u_\infty|}$ of the velocity $u$ are shown in Table \ref{tab:GCL_Error}. As it can be seen, the errors are found to be very small for different grids which demonstrate that the GCL is satisfied in the present model.
\begin{table} \centering
\caption{Relative error $\epsilon$ of the velocity $u$ for the free-stream preservation problem.} 
\label{tab:GCL_Error}
\begin{tabular}{ccl}
\hline \hline
Mesh($\Delta x /L$)     &$\epsilon$         \\ \hline
$0.1$ &  $1.117\times 10^{-15}$                   \\ 
$0.05$    &$4.028\times 10^{-16}$                 \\ 
$0.025$     &$3.872\times 10^{-16}$                \\ \hline \hline
\end{tabular}
\end{table}

\subsection{Flow over NACA0012 airfoil in plunging motion}
We consider a flow over an airfoil in a plunging motion to test the capability of the solver in handling complex vortex-dominated flows. It is known that the flow over plunging airfoil produces thrust over a wide range of oscillation frequencies \citep{zheng2012study}, the phenomenon known as Knoller–Betz effect \citep{jones1998experimental}.
Another interesting phenomenon that has been observed in experiments \citep{jones1998experimental} is the formation of asymmetric deflected wake pattern at high Strouhal numbers, even in symmetrically plunging motions.
Apart from experimental studies \citep{bratt1953flow, jones1998experimental, heathcote2007jet}, several numerical studies \citep{liang2011high, zheng2012study, calderon2014absence, bose2018investigating} have also been done for understanding the physics behind the plunging airfoil and the mechanism of thrust generation.

Here, in order to take into account the effect of compressibility, the numerical setup is identical to the numerical study by \citet{liang2011high} based on the high-order accurate spectral difference ALE solution of the compressible NS equations (SD-NS). A NACA0012 airfoil with chord length $c=200$ is placed in the center of a domain with the size $[40c \times 40c]$. The airfoil is undergoing a sinusoidal plunging motion prescribed as
\begin{align}
    x(t) &= X, \\
    y(t) &= Y - h sin \left(\omega t \right), \label{eq:Plunge_eq}
\end{align}
where $h$ and $\omega$ are plunge motion amplitude and frequency, respectively. The Strouhal number is defined as 
\begin{align}
Sr = h \omega / u_\infty,   
\end{align}
and $u_\infty$ is the free-stream velocity. 

Two different scenarios are considered here: slow plunging and fast plunging motions. For both cases the Reynolds number based on the free-stream velocity $u_\infty$ is set to $Re = \rho_\infty u_\infty c / \mu  = 1850$, the Mach number is Ma $= u_\infty / \sqrt{\gamma T_\infty} = 0.2$ and the free-stream temperature is $T_\infty = 0.3$.

\subsubsection{Slow plunging motion}
\begin{figure}
    \centering
    \includegraphics[width=0.5\textwidth]{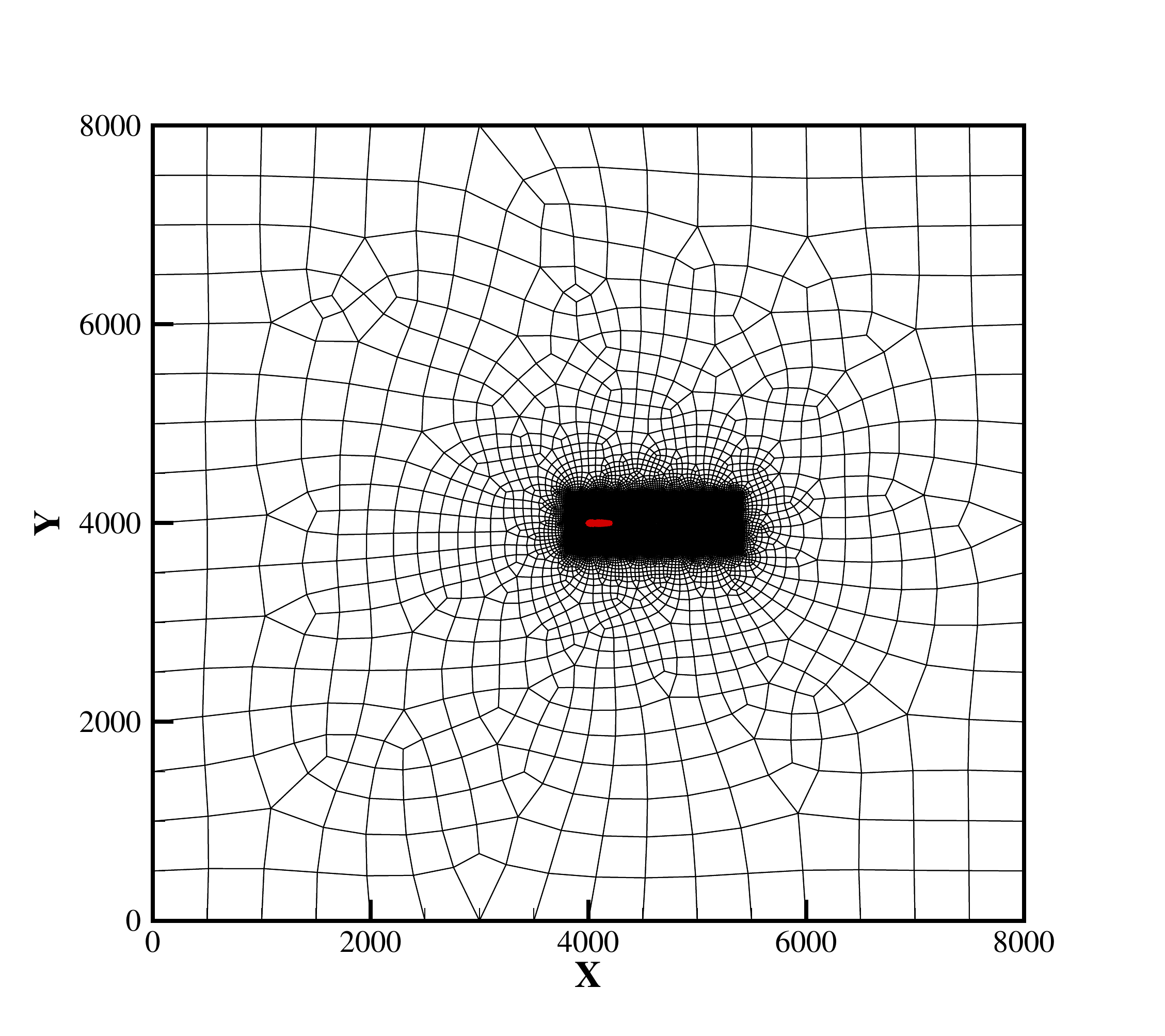}
    \includegraphics[width=0.5\textwidth]{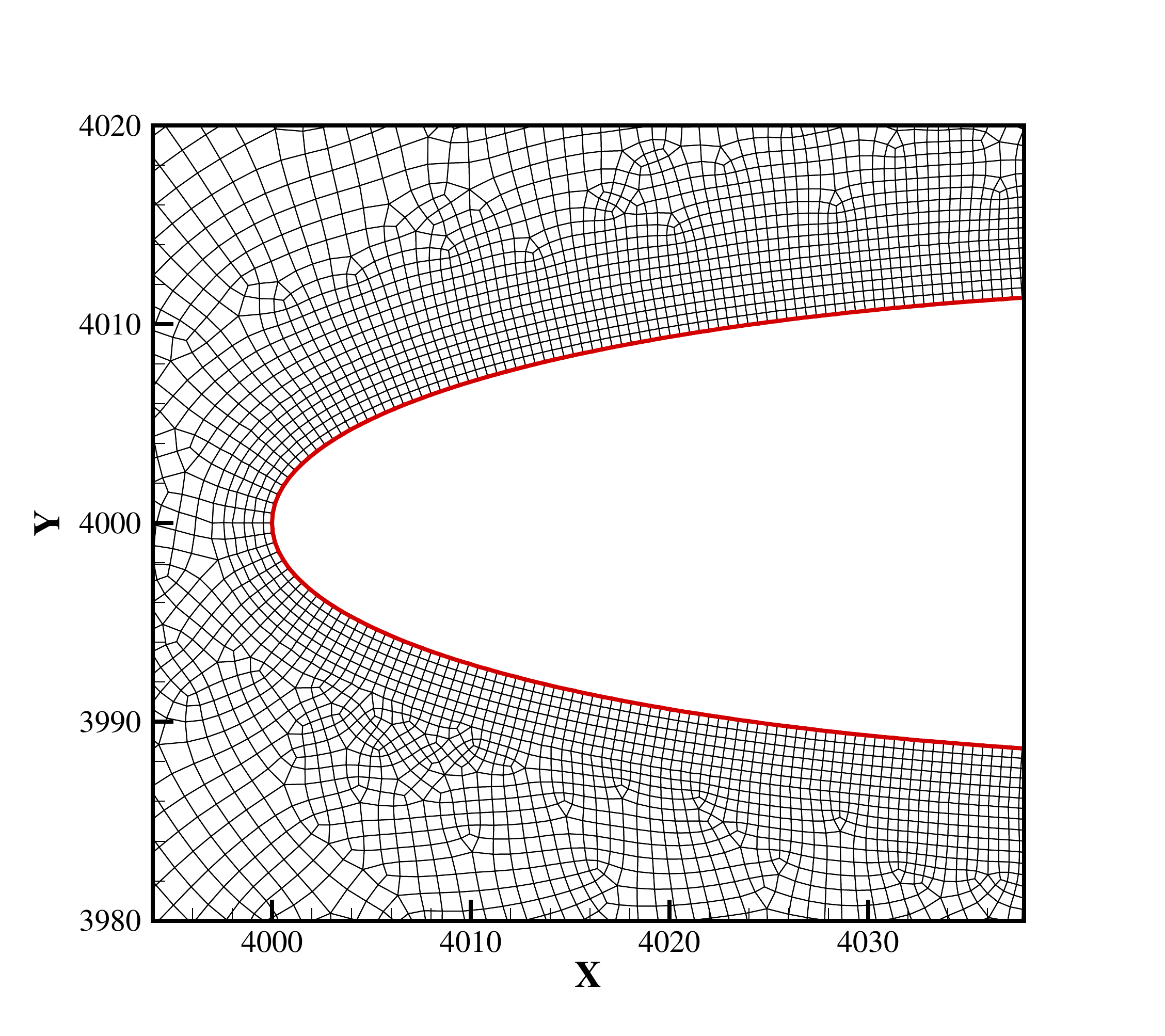}
    \caption{Mesh-2 used for the flow over plunging NACA0012 airfoil. Left: overall view; Right: zoom near leading edge of the airfoil.}
    \label{fig:Airfoil_Grids}
\end{figure}

In the slow plunging motion, the plunge amplitude is $h = 0.08c$ and the Strouhal number is $Sr = 0.46$. We first compute this case using two different meshes with minimum cell sizes of $\delta \approx 0.7$ (Mesh-1) and $\delta \approx 0.5$ (Mesh-2), in order to investigate grid independence. Part of the mesh is shown in Fig. \ref{fig:Airfoil_Grids} , where orthogonal grid is used close to the wall to accurately resolve the boundary layer and anisotropic unstructured gird is used elsewhere. In order to correctly capture the vortical patterns in the wake area, a high resolution mesh with cell size of $\delta \approx 10$ is used in the rectangular domain around the airfoil. Moreover, to minimize the computational cost, the mesh outside of the rectangular domain is highly coarse witch makes the ratio between largest and smallest cell size to be of approximately 1000.  

\begin{figure}
    \centering
    \includegraphics[width=0.5\textwidth]{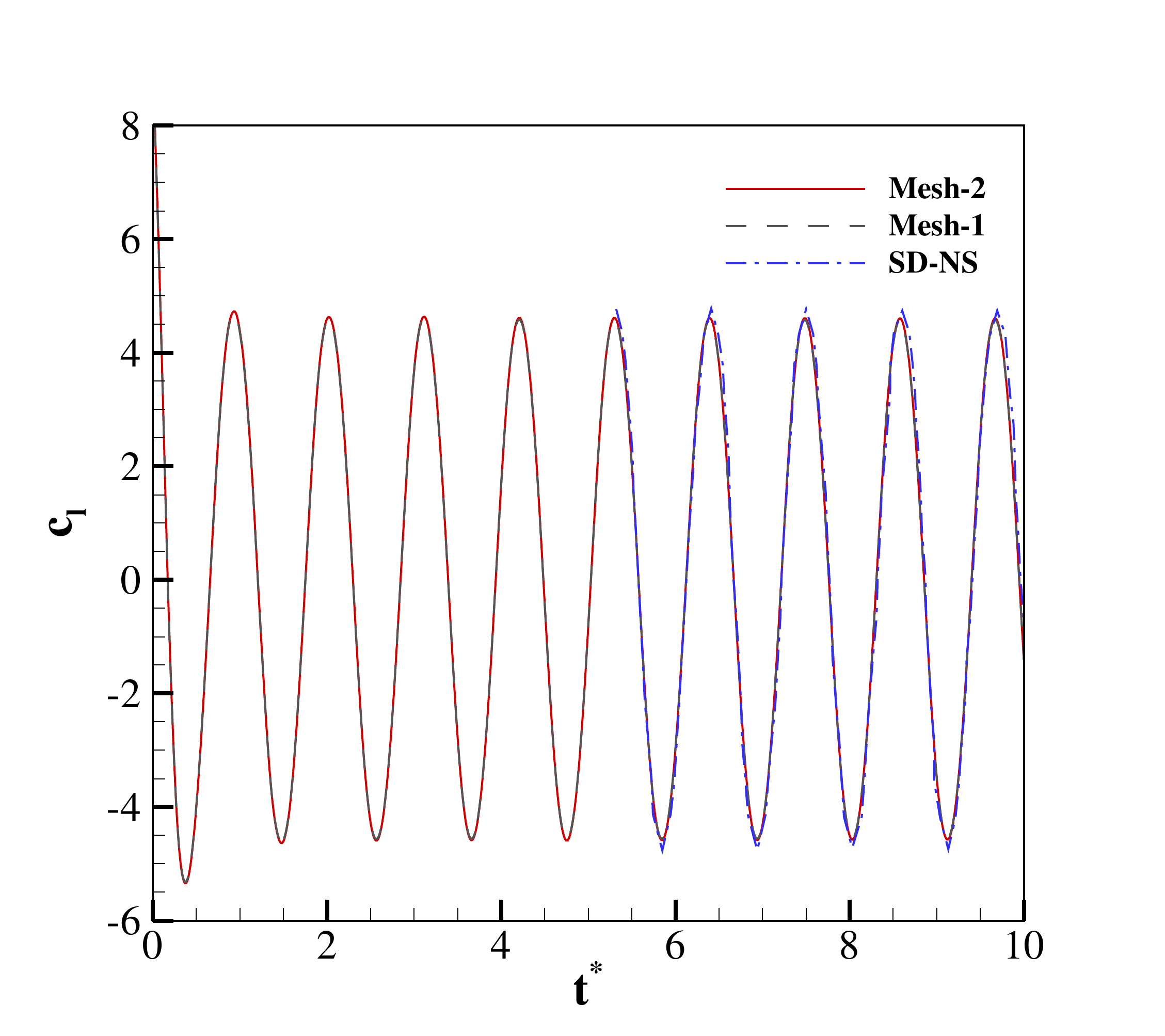}
    \includegraphics[width=0.5\textwidth]{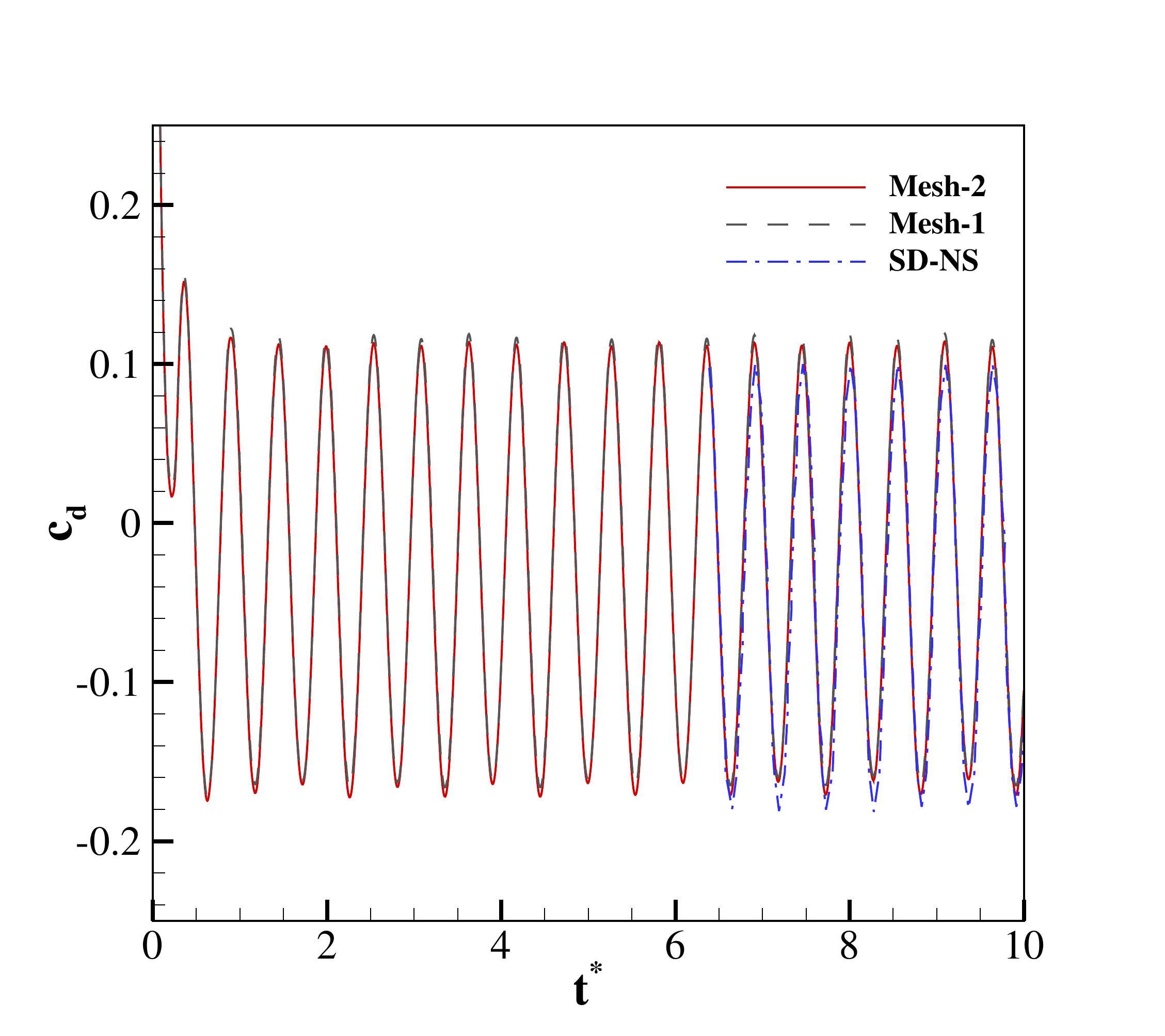}
    \caption{Time evolution of lift (left) and drag (right) coefficients for slow plunging motion of NACA0012 airfoil with $h=0.08c$, $Sr=0.46$ and $Ma=0.2$.}
    \label{fig:Slow_Plunge_Force}
\end{figure}

The time evolution of the aerodynamic forces predicted by both grids are compared in Fig. \ref{fig:Slow_Plunge_Force}. The lift coefficient is defined as $c_L = F_L / (0.5 \rho_\infty u_\infty^2 c)$, where $F_L$ is the total lift force acting on the airfoil and the drag coefficient is given by $c_D = F_D / (0.5 \rho_\infty u_\infty^2 c)$, where $F_D$ denotes the total drag force. As it can be seen in Fig. \ref{fig:Slow_Plunge_Force}, lift coefficient varies symmetrically about zero mean, however, drag coefficient oscillates around a negative average value, which means that a small thrust is generated in this case. Moreover, the two grids give almost identical results which shows convergence to a grid independence solution. 
To validate the solver, the numerical results of \citet{liang2011high} over some cycles are also shown in Fig. \ref{fig:Slow_Plunge_Force}. It is observed that the results are in good agreement. 

\begin{figure}
    \centering
    \includegraphics[width=0.5\textwidth]{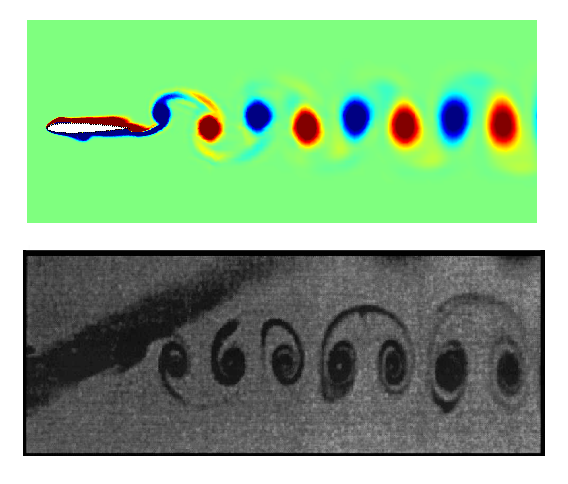}
    \caption{Vorticity computed by the present model (top) and the experimental results reported by \citet{jones1998experimental} (bottom), for slow plunging motion of NACA0012 airfoil with $h=0.08c$ and $Sr=0.46$. Contour levels are bounded between $ -6 \le \Omega c / u_\infty \le 6$.}
    \label{fig:SlowPlung_Vor}
\end{figure}

Figure \ref{fig:SlowPlung_Vor} shows the vorticity contours obtained by the present model in comparison with the experimental results reported by \citet{jones1998experimental}. Due to relatively low Strouhal number in this case, the leading and trailing edges separation results in an almost symmetric flow pattern which is very similar to the experiment. 

\subsubsection{Fast plunging motion}
Next we consider the fast plunging motion of the NACA0012 airfoil which corresponds to a motion with $h = 0.12c$ and $Sr = 1.5$. The computation is performed using the mesh with the minimum cell size of $\delta \approx 0.5$.

\begin{figure}
    \centering
    \includegraphics[width=0.5\textwidth]{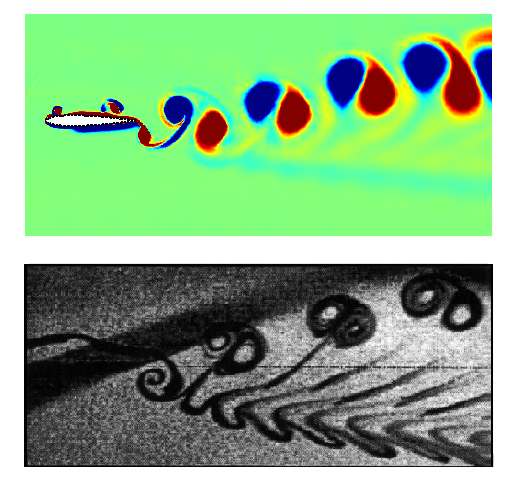}
    \caption{Vorticity computed by the present model (top) and the experimental results reported by \citet{jones1998experimental} (bottom), for fast plunging motion of NACA0012 airfoil with $h = 0.12c$ and $Sr = 1.5$. Contour levels are bounded between $ -6 \le \Omega c / u_\infty \le 6$.}
    \label{fig:Fast_Plunge_Vor}
\end{figure}

Figure \ref{fig:Fast_Plunge_Vor} compares the vorticity contour obtained from the present model with the experimental results of \citet{jones1998experimental}. It can be seen that in this case, the spatial symmetry of the wake vortex pattern is lost and a deflected vortex street is generated. The deflected vortex pattern is travelling upward, because, according to Eq. (\ref{eq:Plunge_eq}), the first stroke is downward. It is also shown that the present model is able to capture dual-mode vortex street, in close resemblance with the experiment. The formation of dynamic stall vortex (DSV) \citep{sangwan2017investigation} near the leading edge of the airfoil is also observed in this figure. Dynamic stall vortices convect towards the trailing edge of the airfoil.  

\begin{figure}
    \centering
    \includegraphics[width=0.5\textwidth]{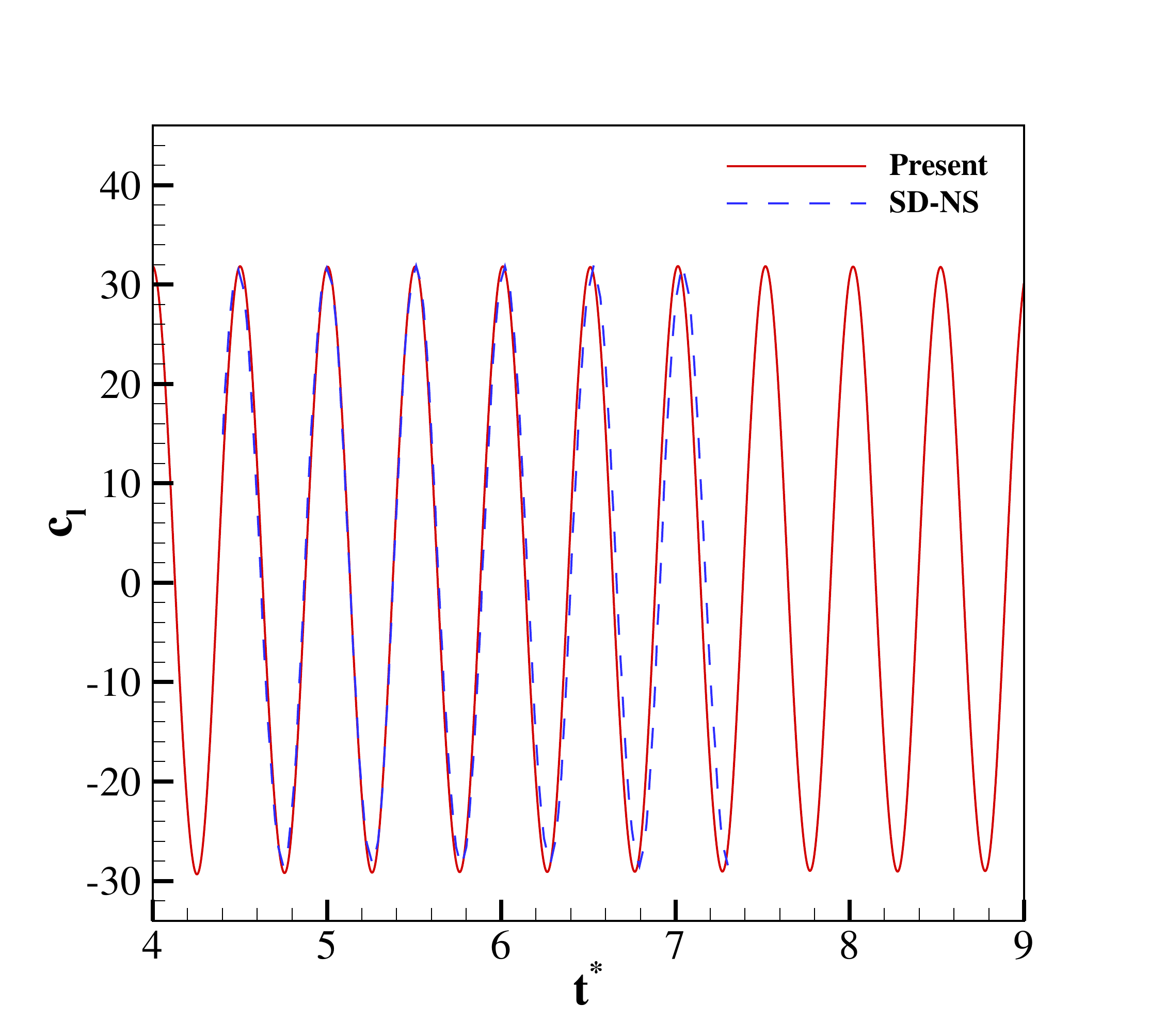}
    \includegraphics[width=0.5\textwidth]{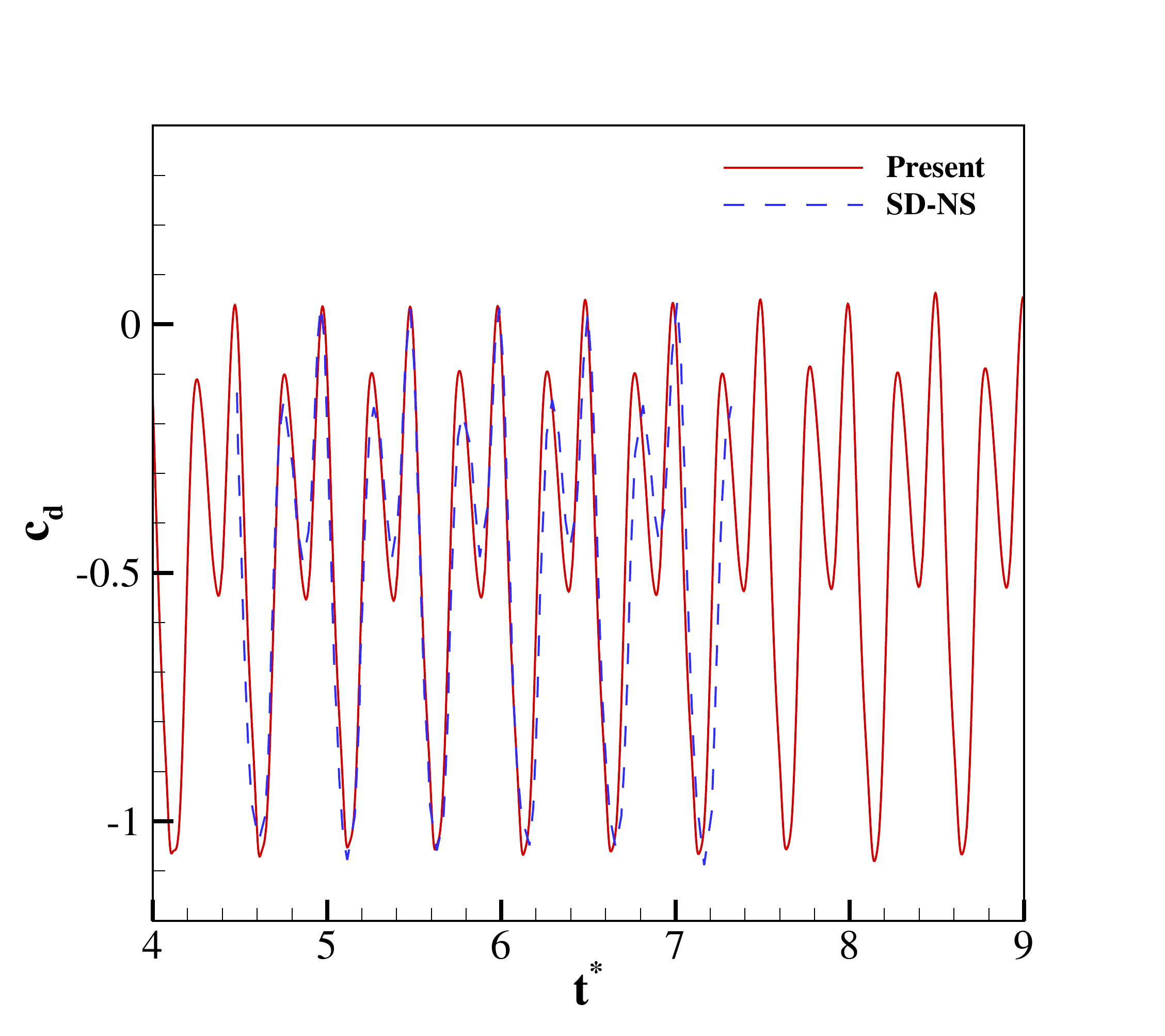}
    \caption{Time evolution of lift (left) and drag (right) coefficients for fast plunging motion of NACA0012 airfoil with $h = 0.12c$, $Sr = 1.5$ and $Ma=0.2$.}
    \label{fig:Fast_Plunge_Force}
\end{figure}

In order to validate the results quantitatively, we compare the time history of the aerodynamic forces computed over several periods by the present model with that of the SD-NS solver \citep{liang2011high}; excellent agreement is observed. As it is shown in Fig. \ref{fig:Fast_Plunge_Force}, the maximum value of lift is larger than the slow plunging case and it oscillates symmetrically around a small mean value of about $1.43$. The drag coefficient, on the other hand, is asymmetric and mainly negative which results in a net mean thrust. 


\subsection{Flow over NACA0012 airfoil in pitching motion}
Now, we turn our attention to a flow over pitching airfoil. The experimental works conducted by \citet{koochesfahani1989vortical},  \citet{bohl2009mtv} and \citet{mackowski2015direct} are among most comprehensive studies on flow over pitching airfoil in the incompressible regime, where they studied the vortical patterns in the wake and measured the thrust coefficient as the function of the reduced frequency. Experiments show that the thrust coefficient increases monotonically with pitching frequency. However, the pure pitching motion is not, in general, an effective mechanism for producing thrust \citep{mackowski2015direct}. There are also several numerical studies in the literature about different aspects of underlying fluid dynamics involved in pitching airfoil \citep{young2004oscillation,liang2011high,bose2018investigating}.

In this case, we consider the flow past NACA0012 airfoil in the pure pitching motion along its quarter chord axis ($c/4$). Therefore, the motion can be prescribed by the following expression
\begin{align}
    x(t) &= (X - X_c) cos(\theta) - (Y - Y_c) sin(\theta), \\
    y(t) &= (X - X_c) sin(\theta) + (Y - Y_c) cos(\theta),
\end{align}
where $(X_C,Y_C)$ is the center of rotation, $\theta = A sin(\omega t)$ is the pitching angle, $A$ denotes the pitch motion amplitude and $\omega$ is pitching frequency. The reduced frequency of pitching is defined as 
\begin{align}
k = \omega c / 2 u_\infty.
\end{align}

First, the Mach number is considered to be $Ma=0.08$ to avoid significant effect of compressibility and to compare the results with the water tunnel experiment data \citep{bohl2009mtv, mackowski2015direct}. The simulation is performed at pitching amplitude of $A = 2^\circ$, reduced frequencies of $k = 0$ (stationary), $k=6.68$ and $k=10$ and at Reynolds number $Re = 12000$. The high Reynolds number makes this test-case more challenging, although the flow is still considered to be laminar. The mesh used for the computations has minimum cell size of $\delta \approx 0.2$ close to the wall. 

\begin{figure}
    \centering
    \includegraphics[width=0.5\textwidth]{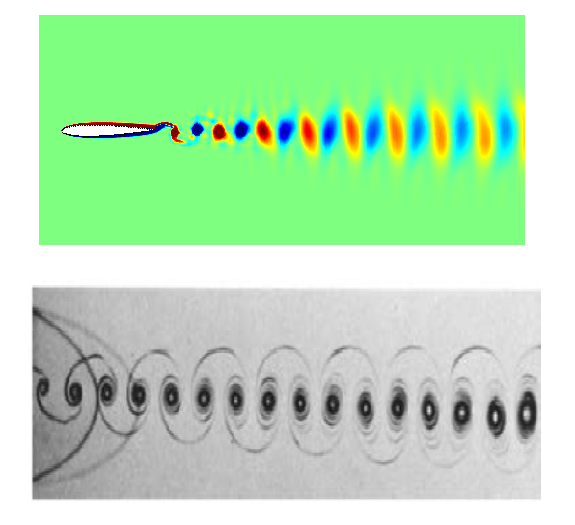}
    \caption{Vorticity computed by the present model (top) and the experimental results reported by \citet{koochesfahani1989vortical}. (bottom), for pitching motion of NACA0012 airfoil with $A = 2^\circ$ and $k=6.68$. Contour levels are bounded between $ -18 \le \Omega c / u_\infty \le 18$.}
    \label{fig:PitchVor_k=668}
\end{figure}

The vortical pattern obtained by the present model is shown in Fig. \ref{fig:PitchVor_k=668} 
in comparison with the experimental results of \citet{koochesfahani1989vortical}, where similar pattern can be observed. The vortex pattern of the present simulation are also quite consistent with other numerical results in the literature \citep{liang2011high, zheng2012study}.

\begin{figure}
    \centering
    \includegraphics[width=0.5\textwidth]{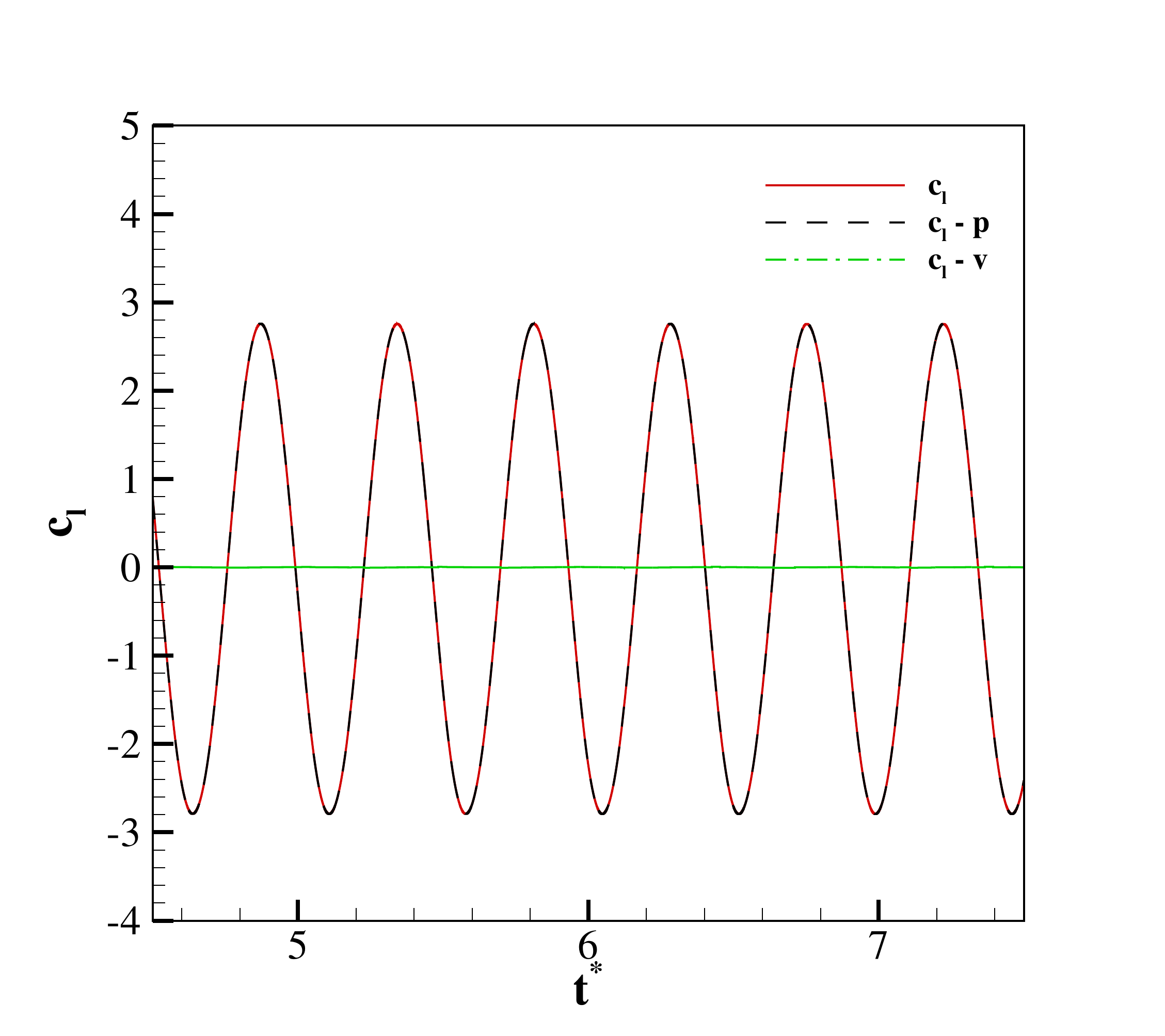}
    \includegraphics[width=0.5\textwidth]{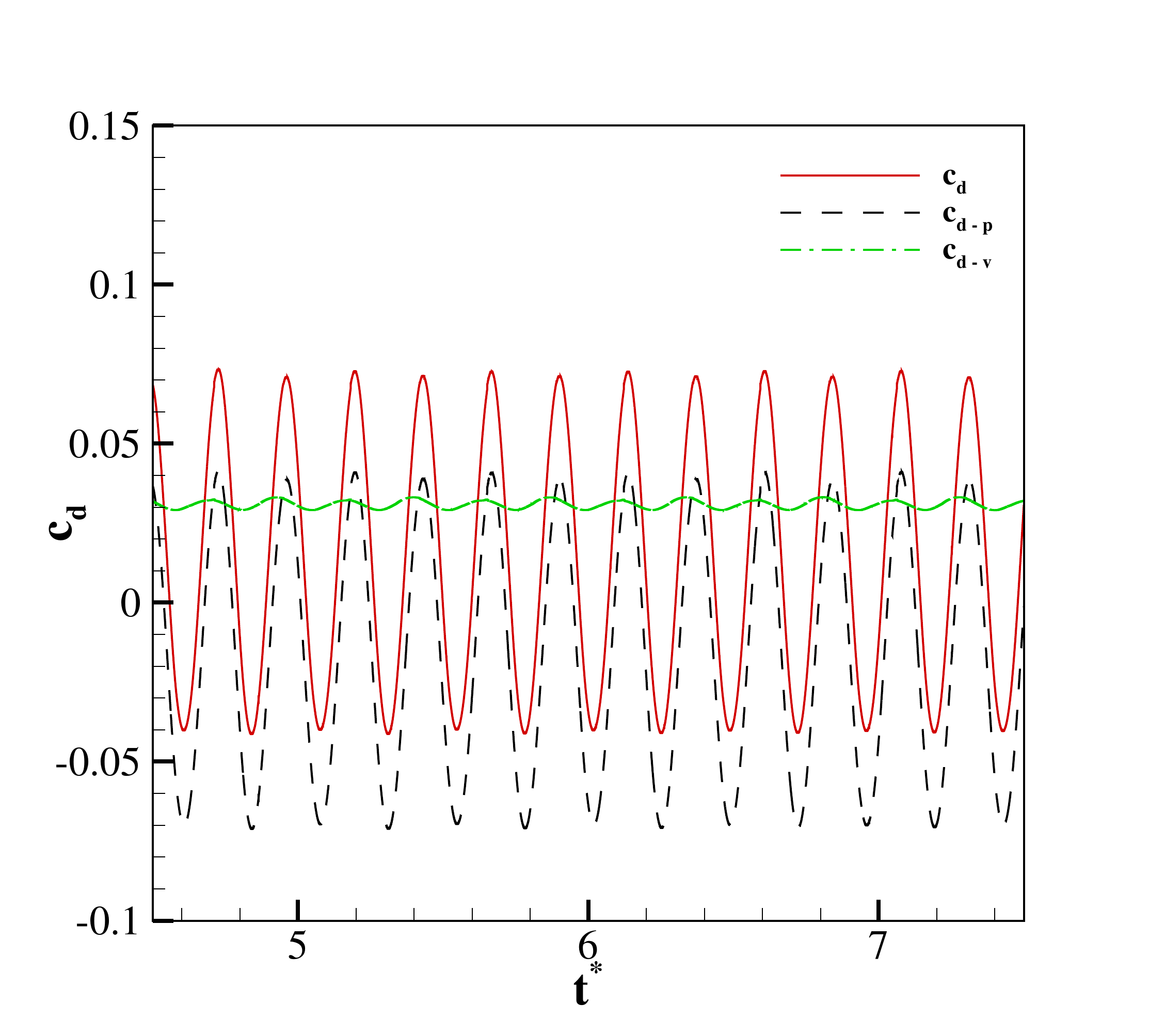}
    \caption{Time evolution of lift (left) and drag (right) coefficients for pitching motion of NACA0012 airfoil with $A = 2^\circ$, $k=6.68$ and $Ma = 0.08$.}
    \label{fig:PitchForce_k=668}
\end{figure}

\begin{figure}
    \centering
    \includegraphics[width=0.5\textwidth]{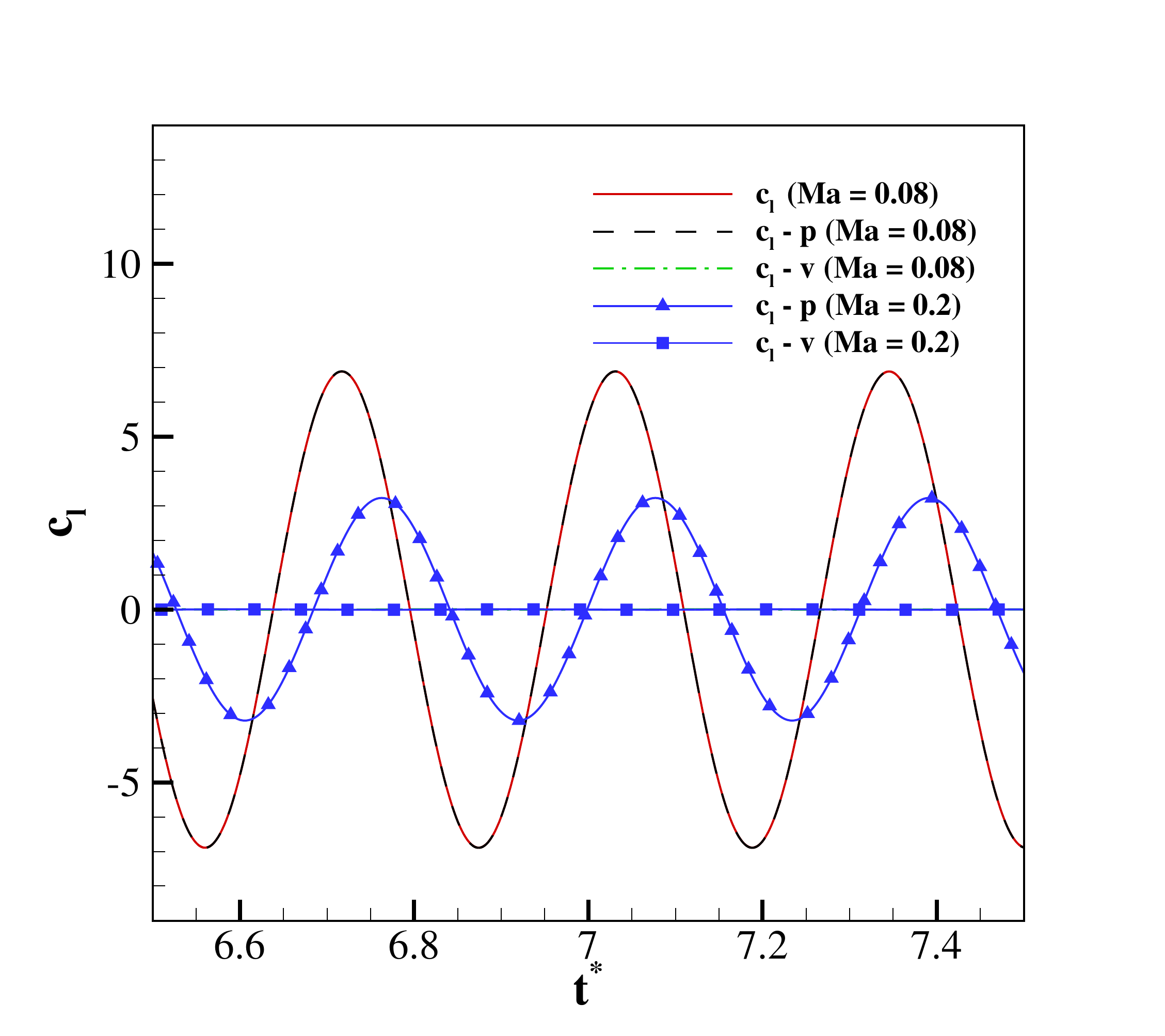}
    \includegraphics[width=0.5\textwidth]{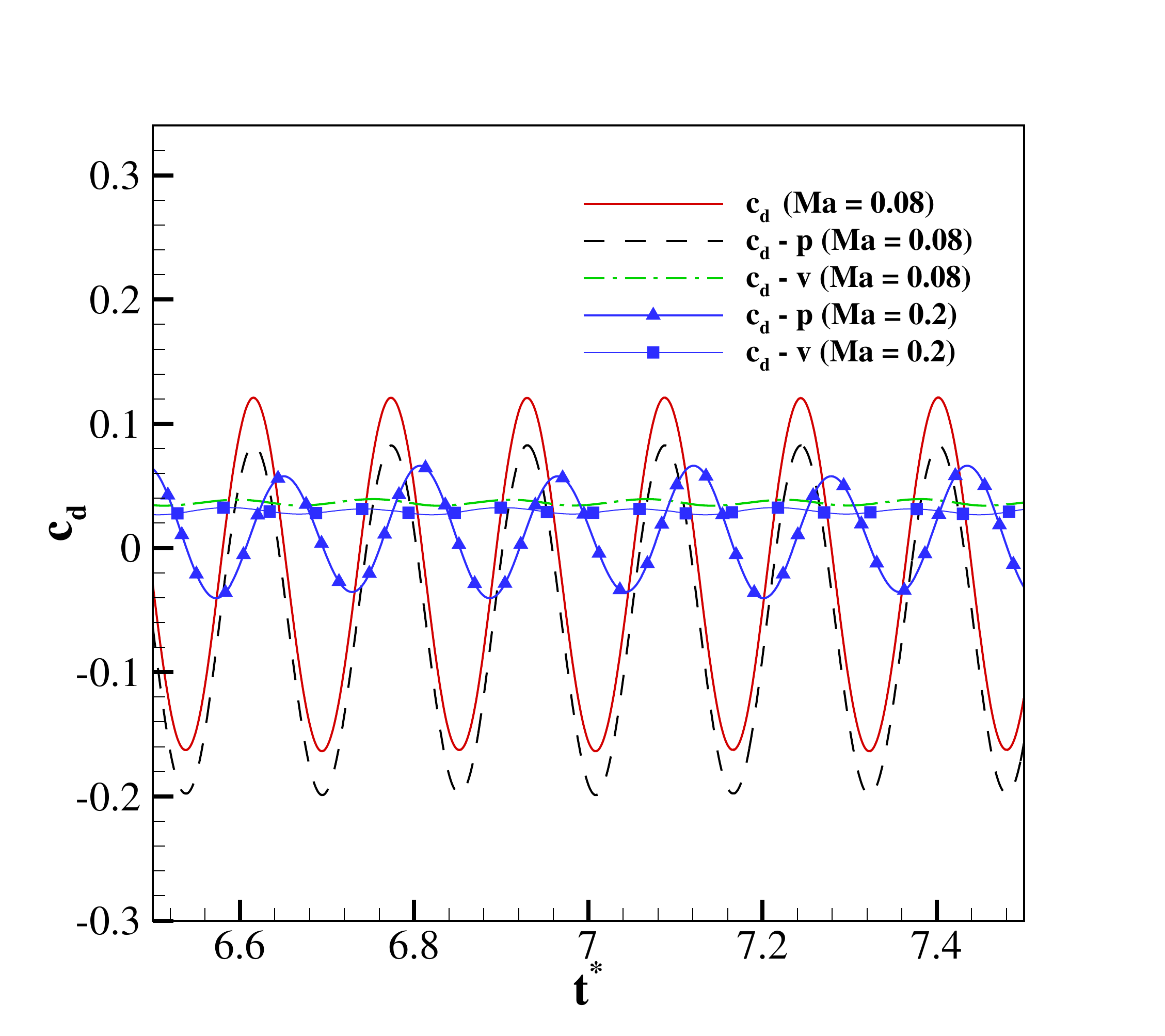}
    \caption{Time evolution of lift (left) and drag (right) coefficients for pitching motion of NACA0012 airfoil with $A = 2^\circ$, $k=10$. Lines: $Ma=0.08$; lines with symbols: $Ma=0.2$.}
    \label{fig:PitchForce_k=10}
\end{figure}

The time history of aerodynamic forces are presented in Figs. \ref{fig:PitchForce_k=668} and \ref{fig:PitchForce_k=10} for reduced frequencies of $k=6.68$ and $k=10$, respectively. In both cases, the lift force acting on the airfoil is only due to pressure term ($c_{l-p}$) while the contribution from the viscous force ($c_{l-v}$) vanishes. Under this condition, the average lift is zero. The drag force however, has contributions from both the pressure ($c_{d-p}$) and the viscous forces ($c_{d-v}$). There is an average drag force acting on the airfoil in the case of $k=6.68$ and a small thrust in the case of $k=10$. 

To investigate the effect of compressibility and in accordance with the numerical simulation of \citet{young2004oscillation} based on the finite-difference discretization of compressible NSF equations, we repeated the simulations at higher Mach number $Ma=0.2$. The time histories of lift and drag coefficients in this case at reduced frequency of $k=10$ are shown in Fig. \ref{fig:PitchForce_k=10} in comparison with the results of the low Mach number case. It can be seen that the compressibilty effect significantly changes the distribution of pressure force, while the viscous force remains almost the same. 

\begin{figure}
    \centering
    \includegraphics[width=0.5\textwidth]{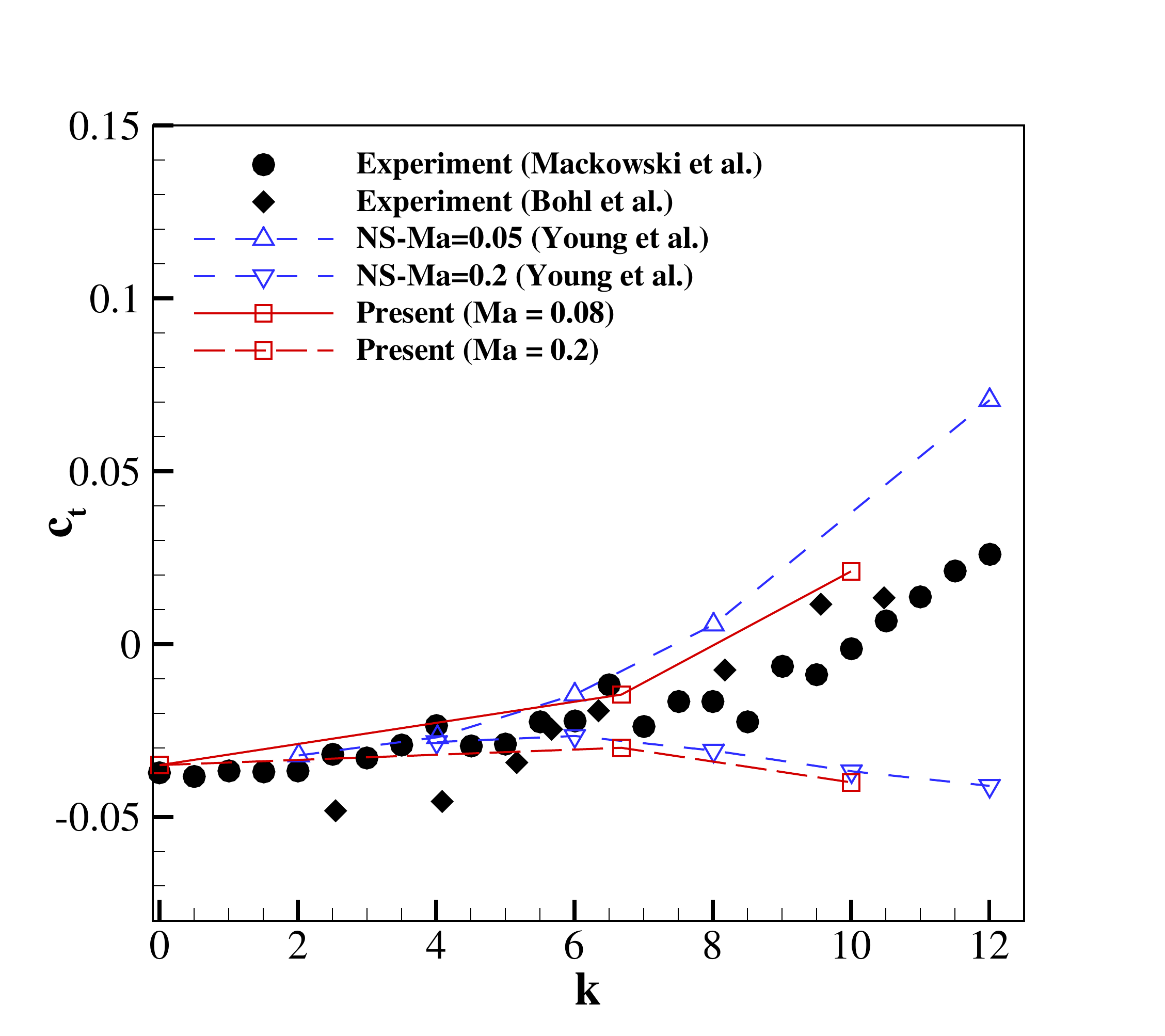}
    \caption{Comparison of the mean thrust coefficient with experimental and numerical results.}
    \label{fig:Pitch_ct}
\end{figure}

Finally, Fig. \ref{fig:Pitch_ct} shows the mean thrust coefficient of the present model at different frequencies in comparison with the experimental and numerical results. In the low Mach number case, the thrust coefficient shows monotonic behaviour with frequency. However, the case with $Ma=0.2$ shows a significantly different behaviour due to the effect of compressibilty. We therefore conclude that pure pitching motion is even less efficient in producing thrust when the flow speed increases and the compressibilty effects become important. 

\subsection{Transonic flow over NACA0012 airfoil in pitching motion}
Finally, we solve a more challenging problem of a transonic flow over NACA0012 airfoil in pitching motion. Accurate computations of unsteady transonic flow is relevant in many applications such as wing flutter analysis or rotor-blade design \citep{chyu1981calculation}.

We set the free-stream Mach number to $Ma=u_\infty/\sqrt{\gamma T_\infty}=0.85$, with $T_\infty = 0.3$, Reynolds number $Re = 10000$, pitching amplitude $A=2^\circ$ and reduced frequency of $k=3$. Due to the high Mach number in this simulation, we need to employ the shifted lattices as presented in \cite{saadat2019lattice}. In our application, we use the lattice with a shift only in the free-stream direction as $\bm{U}=(U_x,U_y)=(0.3,0)$. In this way, deviations in the pertinent higher-order moments are minimized whenever the flow velocity is around $\bm{U}$, and this transformation makes possible to simulate high Mach number flows \citep{frapolli2016lattice}. For further details on the present model with shifted lattices the reader is referred to \citet{saadat2019lattice}. 
\begin{figure}
    \centering
    \includegraphics[width=0.5\textwidth]{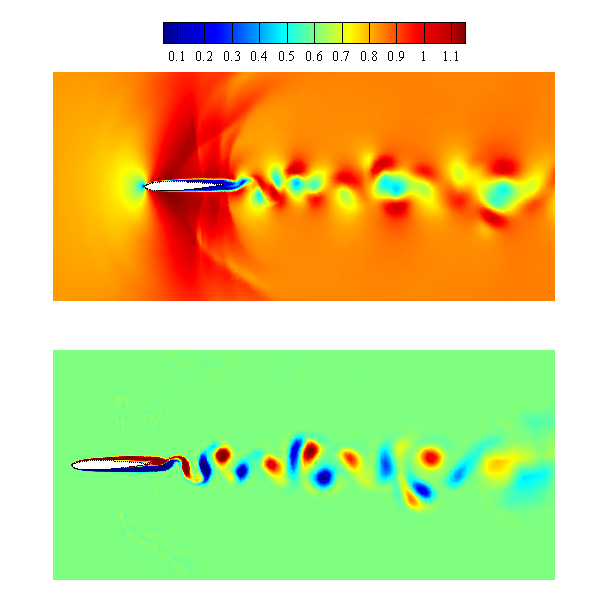}
    \caption{Mach number (top) and Vorticity (bottom) computed by the present model for pitching motion of NACA0012 airfoil with $A = 2^\circ$, $k=3.0$ and $Ma=0.85$. Vorticity contour levels are bounded between $ -11 \le \Omega c / u_\infty \le 11$.}
    \label{fig:Transonic}
\end{figure}

Fig. \ref{fig:Transonic} shows the Mach number distribution over the airfoil and vorticity contours computed by the present model. It is observed that in this case, a complex flow field is formed with multiple shock waves interacting with the boundary and shear layers. Downstream of the leading edge, the flow accelerates causing a formation of weak oblique shock when it reaches the boundary layer. Weak shock waves in the form of lambda-shocks appear further downstream as well. These shock waves interact with the boundary layer, causing the flow separation, and will also be influenced by the vortex shedding downstream of the airfoil \citep{mittal1998finite}. The vortex shedding associated with the shear layer instabilities combines with the vortex shedding due to the airfoil movement, resulting in a complex vortex pattern in the wake region. 
\begin{figure}
    \centering
    \includegraphics[width=0.5\textwidth]{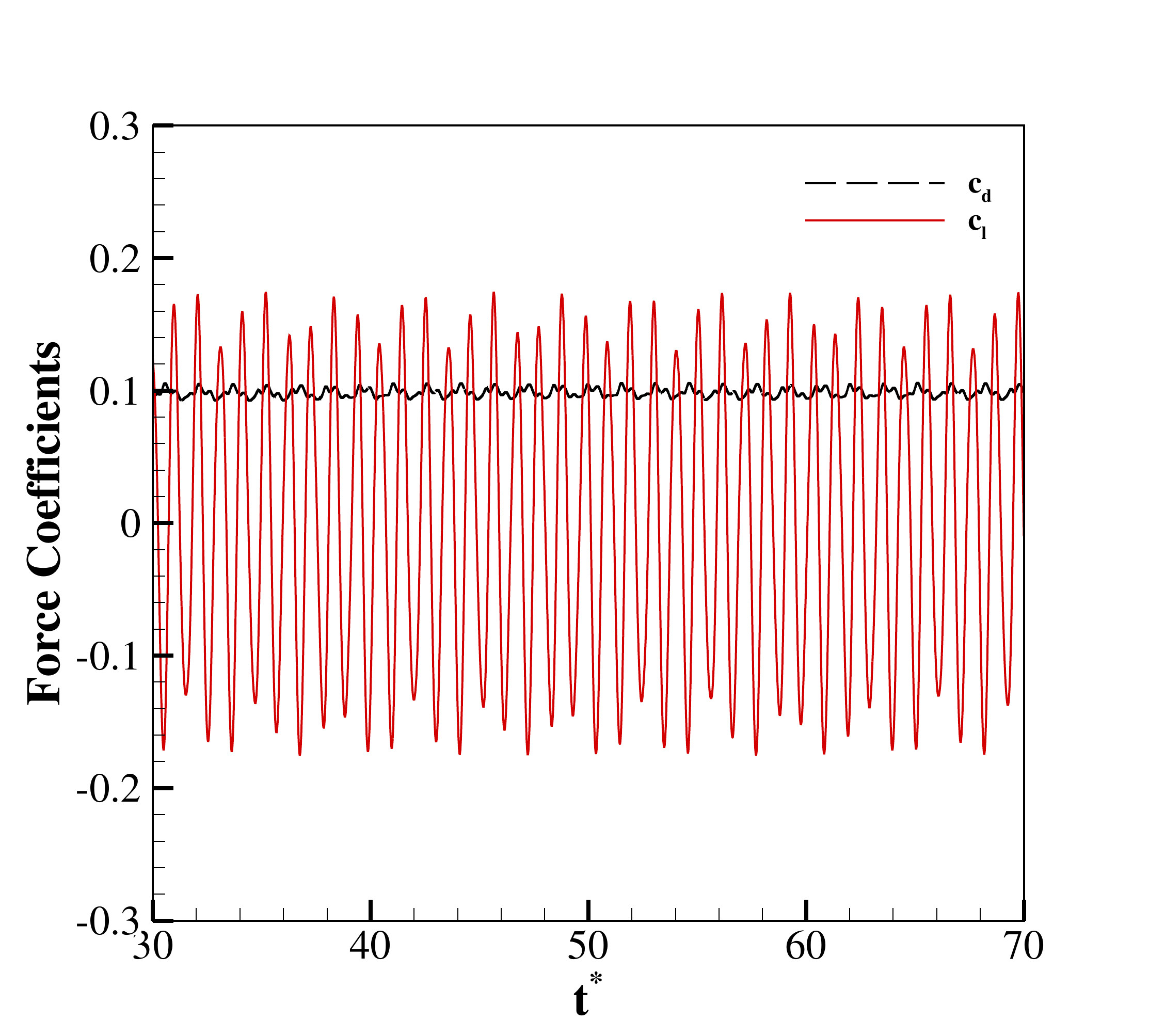}
    \caption{Time evolution of lift and drag coefficients for pitching motion of NACA0012 airfoil with $A = 2^\circ$, $k=3$ and $Ma=0.85$.}
    \label{fig:Transonic_cld}
\end{figure}

Time histories of lift and drag coefficients are presented in Fig. \ref{fig:Transonic_cld}. Similar to the previous pitching cases with smaller Mach number, the average lift force is close to zero. However, there is a mean drag force of $c_d \approx 0.0986$ acting on the airfoil, which is significantly larger than its counterpart in low Mach number case. This confirms the previous observation about the effect of compressibilty on increasing the drag force.

\section{Concluding remarks} \label{Sec:conclusion}
In this work, we proposed a solution methodology for the simulation of compressible flows on unstructured moving meshes based on the arbitrary Lagranian-Eulerian (ALE) technique applied to a compressible lattice Boltzmann model. The kinetic equations of the compressible LB model on standard lattices were first mapped from a physical moving domain to a fixed computational domain. The resulting equations were solved by employing the second-order accurate finite element interpolation. It was shown, both theoretically and numerically, that the problem regarding the geometric conservation law (GCL), which needs special treatments in the ALE-Navier-Stokes solvers does not appear here and the proposed ALE-LB model satisfies the GCL condition exactly. The analysis of the model was conducted through simulation of compressible flow over NACA0012 airfoil undergoing plunging and pitching motions at different Mach numbers. 

Most of the LB models in the literature employ a fixed background Cartesian grid, however, the present model is based on the body-fitted unstructured mesh which is more efficient in resolving small scale flow structures near the wall. Moreover, unlike previous LB studies, which were limited to low-speed incompressible flow, the LB model considered here is a compressible model which covers the range from subsonic to moderately supersonic regimes.

It was demonstrated that the model is able to properly predict the relevant features of the complex flow over flapping airfoil. In particular, the vortical patterns of wake, the time histories of lift and drag coefficients and their mean values agreed well with the experimental and numerical results in the literature. Both slow and fast plunging motion of airfoil produce a net mean thrust with very small average lift. Pitching motion, however, is not as effective as plunge motion and a thrust is generated at higher frequencies, and only when the compressibilty effects are small. It was also observed that the impact of compressibilty is mainly on the distribution of pressure force rather than the viscous force. Finally, in order to show the model's performance in simulating high-speed flows, transonic flow over pitching airfoil was considered, where complex flow pattern involving multiple shock waves interacting with the boundary and shear layers were observed in the flow field.

The promising results of the proposed model open interesting prospects toward the numerical simulation of more complex flows such as the dynamic stall problem in compressible flows, flows involving multiple moving/deforming objects or fluid-solid interaction (FSI) problems. For problems including deformation or relative motion of multiple objects, a blending function is needed to construct the mapping function, as it was proposed in \citet{persson2009discontinuous}. This would be the focus of our future research. Extension of the methodology to three dimensions is also another subject of our future works.

\begin{acknowledgments}
This work was supported by the ETH research grant ETH-13 17-1 and the European Research Council (ERC) Advanced Grant No. 834763-PonD. The computational resources at the Swiss National Super Computing Center CSCS were provided under the grant s897.\\
The authors would like to thank Fabian B{\"o}sch and Benedikt Dorschner for useful discussions.
\end{acknowledgments}


%

\end{document}